\documentclass[12pt,a4paper]{article}
\usepackage{amsmath}
\usepackage{amsthm}
\usepackage{graphicx}
\allowdisplaybreaks[1]
\newcommand{\os}{\underline}
\newcommand{\mc}{\multicolumn}
\newcommand{\pr}{\rightarrow}

\newcommand{\xal}{x^{\alpha}}

\newcommand{\cak}{{\cal K}}
\newcommand{\caf}{{\cal F}}
\newcommand{\cax}{{\cal X}}
\newcommand{\cael}{{\cal L}}
\newcommand{\ba}{\begin{array}}
\newcommand{\ea}{\end{array}}

\newcommand{\varp}{\varphi}
\newcommand{\eps}{\varepsilon}

\newcommand{\rln}{{\rm ln}}

\newcommand{\il}{\int\limits}

\newenvironment{inspring}[1]%
{\begin{list}{}{\setlength{\rightmargin}{0cm}
                \setlength{\listparindent}{0cm}
                \settowidth{\labelwidth}{\mbox{#1}}
                \setlength{\leftmargin}{1.1\labelwidth}
                \setlength{\labelsep}{.1\labelwidth}}}%
{\end{list}}

\newcommand{\bi}[1]{\begin{inspring}{#1}}
\newcommand{\ei}{\end{inspring}}

\newcommand{\beq}{\begin{equation}}
\newcommand{\eq}{\end{equation}}
\catcode`\@=11

\font\tenmsa=msam10 \font\sevenmsa=msam7 \font\fivemsa=msam5
\font\tenmsb=msbm10 \font\sevenmsb=msbm7 \font\fivemsb=msbm5
\newfam\msafam
\newfam\msbfam
\textfont\msafam=\tenmsa  \scriptfont\msafam=\sevenmsa
  \scriptscriptfont\msafam=\fivemsa
\textfont\msbfam=\tenmsb  \scriptfont\msbfam=\sevenmsb
  \scriptscriptfont\msbfam=\fivemsb

\def\Bbb{\ifmmode\let\next\Bbb@\else
 \def\next{\errmessage{Use \string\Bbb\space only in math mode}}\fi\next}
\def\Bbb@#1{{\Bbb@@{#1}}}
\def\Bbb@@#1{\fam\msbfam#1}

\newcommand{\dE}{{\Bbb E}}

\newtheorem{thm}{Theorem}
\newtheorem{lem}[thm]{Lemma}

\numberwithin{thm}{section}

\title{Analysis of quantities determining the critical inverse temperature in the annealed Potts model with Pareto vertex weights}
\author{A.J.E.M.\ Janssen \\
Eindhoven University of Technology \\
Department of Mathematics and Computer Science}
\date{}

\begin{document}
\maketitle
\mbox{} \\ \\ \\ \\ \\
\noindent
{\bf Abstract.} \\
We consider in this work the crucial quantity $t_c$ that determines the critical inverse temperature $\beta_c$ in the $q$-state Potts model on sparse rank-1 random graphs where the vertices are equipped with a Pareto weight density $(\tau-1)\,w^{-\tau}\,\cax_{[1,\infty)}(w)$. It is shown in \cite{ref1} that this $t_c$ is the unique positive zero of a function $\cak$ that is obtained by an appropriate combination of the stationarity condition and the criticality condition for the case the external field $B$ equals 0 and that $q\geq3$ and $\tau\geq4$, see \cite{ref1}, Theorem~1.14 and Theorem ~1.21 and their proofs in \cite{ref1}, Section~7.1 and Section~7.3. From the proof of \cite{ref1}, Theorem~1.14, it is seen that $\cak'$ and $\cak''$ also have a unique positive zero, $t_c'$ and $t_c''$, respectively, and $t_c'=t_b$ and $t_c''=t_{\ast}$, where $t_b$ and $t_{\ast}$ are the unique positive zeros of $\caf_0(t)-t\,\caf_0'(t)$ and $\caf_0''(t)$, respectively. Here, $\caf_0(t)=\dE\,[W(e^{tW}-1)/(\dE\,[W]\,(e^{tW}+q-1))]$, and $t_c$, $t_b$ and $t_{\ast}$ play a key role in the graphical analysis of \cite{ref1}, Section~5.1 and Figure~1. Furthermore, $\gamma_c=\exp(\beta_c)-1$ and $t_c$ are related according to $\gamma_c=t_c/\caf_0(t_c)$.

We analyse $t_c$, $t_c'$ and $t_c''$ for general real $\tau\geq4$ and general real $q>2$ by an appropriate formulation of their defining equations $\cak(t_c)=\cak'(t_c')=\cak''(t_c'')=0$. Thus we find, along with the inequality $0<t_c''<t_c'<t_c<\infty$, the simple upper bounds $t_c<2\,{\rm ln}(q-1)$, $t_c'<\frac32\,{\rm ln}(q-1)$, $t_c''<{\rm ln}(q-1)$, as well as certain sharpenings of these simple bounds and counterparts about the large-$q$ behaviour of $t_c$, $t_c'$ and $t_c''$. We show that these (sharpened) bounds are sharp in the sense that they hold with equality for the limiting case $\tau\pr\infty$ (homogeneous case). These results have consequences for $\gamma_c$ and $\beta_c$ via the equation $\gamma_c=t_c/\caf_0(t_c)$. In particular, we find an expression for $\gamma_c$ involving only elementary functions of $t_c$ (without integrals). While the large-$q$ behaviour of $t_c$, $t_c'$ and $t_c''$ is qualitatively the same for all $\tau\geq4$, such a thing does not hold for the behaviour when $q\downarrow2$. It appears that one has to distinguish between the cases $\tau=4$, $4<\tau<5$, $\tau=5$ and $\tau>5$ with decay behaviour as $q\downarrow2$ ranging from exponentially small in $q-2$ to linear in $q-2$. 

\section{Introduction} \label{sec1}
\mbox{} \\[-9mm]

In \cite{ref1} a comprehensive study has been made of the annealed ferromagnetic $q$-state Potts model on sparse rank-1 random graphs, where vertices are equipped with a vertex weight and where the probability of an edge is proportional to the product of the weights. Under a rather general condition on the weight distribution, phase transition is first order, with a unique discontinuity of the order parameter, see \cite{ref1}, Abstract and Section~1. Let $\beta>0$ denote the inverse temperature, and let $\gamma=e^{\beta}-1$ (in \cite{ref1} one writes $\beta'$ instead of $\gamma$, but in the present work the prime $'$ has been reserved to indicate differentiation). Assume that the external field $B$ equals 0 and that $\dE\,[W^2]<\infty$ ($W$ is the random weight variable). Consider the function
\beq \label{e1}
\caf_0(t)=\dE\,\Bigl[\frac{W}{\dE\,[W]}~\frac{e^{tW}-1}{e^{tW}+q-1}\Bigr]~,~~~~~~ t\geq0~.
\eq
The function $\caf_0(t)$ is non-negative, strictly increasing and $\caf_0(t)\pr1$, $t\pr\infty$. Assume that $\caf_0''(t)$ is first positive and then negative, so that $\caf_0(t)$ has a unique inflection point $t_{\ast}>0$ with $\caf_0''(t_{\ast})=0$ and $\caf_0''(t)>0$, $0\leq t<t_{\ast}$, and $\caf_0''(t)<0$, $t>t_{\ast}$. Also consider the annealed pressure
\beq \label{e2}
p(t,\gamma)=\dE\,[{\rm ln}(e^{tW}+q-1)]-\dE\,[W]\,\frac{\gamma}{2q}\, \Bigl((q-1)\Bigl(\frac{t}{\gamma}\Bigr)^2+2\,\frac{t}{\gamma}-1\Bigr)~.
\eq
Under the above condition on $\caf_0$, there is a unique pair $(t_c,\gamma_c)$ with $t_c>0$, $\gamma_c>0$ such that simultaneously the stationarity condition
\beq \label{e3}
\caf_0(t)=\frac{t}{\gamma}
\eq
and the criticality condition
\beq \label{e4}
p(t,\gamma)=p(0,\gamma)
\eq
is satisfied by $(t,\gamma)=(t_c,\gamma_c)$. The criticality condition can be written as
\beq \label{e5}
\frac{1}{\dE\,[W]}\,\dE\,\Bigl[{\rm ln}\Bigl(\frac{e^{tW}+q-1}{q}\Bigr)\Bigr]- \frac{q-1}{2q}\,t\cdot\frac{t}{\gamma}-\frac{t}{q}=0~.
\eq
In Theorem~1.14 of \cite{ref1} about the critical inverse temperature $\beta_c={\rm ln}(1+\gamma_c)$, there is considered the function
\beq \label{e6}
\cak(t)=\frac{1}{\dE\,[W]}\,\dE\,\Bigl[{\rm ln}\Bigl(\frac{e^{tW}+q-1}{q}\Bigr)\Bigr] -\frac{q-1}{2q}\,t\,\caf_0(t)-\frac{t}{q}~,~~~~~~ t\geq0~,
\eq
so that $\gamma$ is eliminated from the left-hand side of (\ref{e5}) by using (\ref{e3}). By \cite{ref1}, Theorem~1.14, $t_c$ is the unique $t>0$ such that $\cak(t)=0$, and
\beq \label{e7}
\beta_c={\rm ln}\Bigl(1+\frac{t_c}{\caf_0(t_c)}\Bigr)~.
\eq
The proof of \cite{ref1}, Theorem~1.14 furthermore shows that
\beq \label{e8}
\cak'(t)=\frac{q-1}{2q}\,[\caf_0(t)-t\,\caf_0'(t)]~,~~~~~~t\geq0~,
\eq
\beq \label{e9}
\cak''(t)={-}\,\frac{q-1}{2q} \,t\,\caf_0''(t)~,~~~~~~t\geq0~.
\eq
According to \cite{ref1}, Section~5.1, for the case $B=0$, there are unique numbers $t_{\ast}$, $t_b>0$ such that
\beq \label{e10}
\caf_0''(t_{\ast})=0~,~~~~~~\caf_0(t_b)-t_b\,\caf_0'(t_b)=0~.
\eq
Denoting the unique positive zeros of $\cak$, $\cak'$ and $\cak''$ by $t_c$, $t_c'$ and $t_c''$ respectively, we thus see that
\beq \label{e11}
t_c=t_c\,,~~t_c'=t_b\,,~~t_c''=t_{\ast}~;~~~~~~0<t_c''<t_c'<t_c<\infty~,
\eq
where we also refer to \cite{ref1}, Figure~1. Hence, $t_c''=t_{\ast}$ and $t_c'=t_b$ have a geometric meaning in terms of the graph of $\caf_0$ (inflection point and point determining the non-trivial tangent line that passes through the origin, respectively). We can find for $t_c$ such a geometric meaning as well from (\ref{e8}). Integrating (\ref{e8}) from 0 to $t$, using $\cak(0)=0$ and partial integration, we get
\beq \label{e11+1}
\cak(t)=\frac{q-1}{q}\,\left(\il_0^t\,\caf_0(s)\,ds-\tfrac12\,t\,\caf_0(t)\right)~,~~~~~~ t\geq0~.
\eq
Hence, the condition $\cak(t_c)=0$ means that
\beq \label{e11+2}
\il_0^{t_c}\,\caf_0(s)\,ds=\tfrac12\,t_c\,\caf_0(t_c)= \il_0^{t_c}\,s\,\caf_0(t_c)/t_c\,ds=\il_0^{t_c}\,s/\gamma_c\,ds~,
\eq
where we recall from (\ref{e3}) that $\gamma_c=t_c/\caf_0(t_c)$. Therefore, $\il_0^{t_c}\,(\caf_0(s)-s/\gamma_c)\,ds=0$, and so the sets $\{(s,v)\,|\,0\leq s\leq t_c,~\caf_0(s)\leq v\leq s/\gamma_c\}$ and $\{(s,v)\,|\,0\leq s\leq t_c,~s/\gamma_c\le v\leq\caf_0(s)\}$ have equal area. We illustrate this in Figure~1 below, where we display for the homogeneous case $W\equiv 1$ with $q=100$ the graph of $\caf_0(t)$, $0\leq t\leq10$ (this example was suggested by C.\ Giardin\`a and C.\ Giberti). In this case, see Section~\ref{sec8}, we have $\caf_0(t)=1-100/(e^t+99)$, and $t_c''=\rln(q-1)=4.5951...\,$, $t_c=2\,\rln(q-1)=9.1902...\,$, while $t_c'$ can be determined numerically as $1.3648...\rln(q-1)=6.2716...\,$. In this Figure~1, we also have indicated on the $t$-axis the points $t_c''$, $t_c'$ and $t_c$, as well as the tangent line passing through $(0,0)$ and $(t_c',\caf_0(t_c'))$ and the line $\{s/\gamma_c\,|\,0\leq s\leq t_c\}$ that nicely realizes the equal area condition (compare \cite{ref1}, Proposition~5.6). In Figure~1 of \cite{ref3}, the equal area condition is illustrated for the case that $W$ has the density $\chi_{[0,1)}(w)$, $w\geq0$, and $q=100$.

\begin{figure}[ht]
\begin{center}
\includegraphics[width=0.6\textwidth]{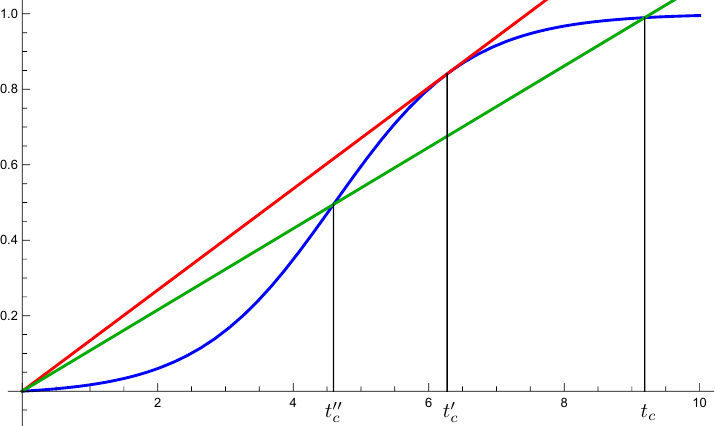}
\end{center}
\caption{Plot of $\caf_0(t)=1-100/(e^t+99)$, $0\leq t\leq10$, for the homogeneous case $W\equiv 1$ and $q=100$ (blue line). The points $t_c''=t_{\ast}$, $t_c'=t_b$ and $t_c$ are determined from the conditions (\ref{e10}) and (\ref{e11+2}) as ${\rm ln}\,99$, $1.3648...{\rm ln}\,99$ and $2\,{\rm ln}\,99$, respectively. The red line is the tangent line to the graph of $\caf_0$ through the point $(t_c',\caf_0(t_c'))$ and the green line is the straight line through $(0,0)$ and $(t_c,\caf_0(t_c))$, exhibiting the equal area condition.}
\label{fig:fig1}
\end{figure}

We have $\cak(0)=\cak'(0)=\cak''(0)=0$ while $\cak(t)$ is negative for small positive $t$ and positive for large positive $t$. Therefore, $\cak$, $\cak'$ and $\cak''$ are first negative and then positive, and for all $t>0$ we have
\beq \label{e12}
t<t_c\Leftrightarrow \cak(t)<0~;~~~~~~t>t_c\Leftrightarrow \cak(t)>0~,
\eq
\beq \label{e13}
t<t_c'\Leftrightarrow\cak'(t)<0~;~~~~~~t>t_c'\Leftrightarrow\cak'(t)>0~,
\eq
\mbox{} \\[-7.5mm]
\beq \label{e14}
t<t_c''\Leftrightarrow\cak''(t)<0~;~~~~~~t>t_c''\Leftrightarrow\cak''(t)>0~.
\eq

\section{Main results} \label{sec2}
\mbox{} \\[-9mm]

We are interested in bounds and approximations of $t_c$, $t_c'$ and $t_c''$ in the particular case that $W$ has Pareto density
\beq \label{e15}
f_W(w)=(\tau-1)\,w^{-\tau}\,\cax_{[1,\infty)}(w)~,~~~~~~w\geq0,
\eq
with $\tau\geq4$. Here $\cax_{[1,\infty)}(w)$ equals 1 for $1\leq w<\infty$ and 0 for $0\leq w<1$. The restriction $\tau\geq4$ ensures that $\caf_0$ in (\ref{e1}) and $\cak$ in (\ref{e6}) have for all $q>2$ a unique, positive inflection point $t_{\ast}=t_c''$, see \cite{ref1}, Theorem~1.21. In \cite{ref2}, a similar effort has been done for $t_c$ in the case that $W$ has an exponential density $f_W(w)=\exp({-}w)\,\cax_{[0,\infty)}(w)$, $w\geq0$. It appears that the results we obtain in the present work are, compared with those obtained in \cite{ref2}, more transparent, complete and appealing and hardly require numerical efforts.

In Section~\ref{sec3}, we shall show that for $t\geq0$, $q>2$, $\tau\geq4$
\beq \label{e16}
\cak(t)=\frac{\tau-2}{\tau-1}\,{\rm ln}\Bigl(\frac{e^t+q-1}{q}\Bigr) +\Bigl(\frac{1}{\tau-1}-\frac{q+1}{2q}\Bigr)t+\frac{(\tau-2)(\tau-3)}{2(\tau-1)}\, t(q-1)\,D~,
\eq
\beq \label{e17}
\cak'(t)=\tfrac12\,(q-1)\Bigl(\frac1q-\frac{\tau-2}{e^t+q-1}+(\tau-2)(\tau-3)\,D \Bigr)~,
\eq
\beq \label{e18}
\cak''(t)=\frac{(q-1)(\tau-2)}{2t}\,\Bigl(\frac{t\,e^t}{(e^t+q-1)^2}-\frac{\tau-3} {e^t+q-1}+(\tau-2)(\tau-3)\,D\Bigr)~,
\eq
where
\beq \label{e19}
D=D(t)=\il_1^{\infty}\,\frac{w^{-\tau+1}\,dw}{e^{tw}+q-1}~.
\eq
It happens to be a very convenient fact that in (\ref{e16}), (\ref{e17}) and (\ref{e18}) the same integral $D$ appears.

In \cite{ref3}, the case that $W$ has the density $\chi_{[0,1)}(w)$, $q\geq0$ has been considered in all detail. It appears that the corresponding functions $\cak$, $\cak'$ and $\cak''$, as well as $\caf_0$, can be expressed, pretty much as in (\ref{e16}), (\ref{e17}), (\ref{e18}) and (\ref{e35}), in terms of elementary functions and one and the same integral of the same type as $D$ in (\ref{e19}). This latter integral can be expressed in terms of the Euler dilogarithm, resulting into sharp bounds for $t_c$, $t_c'$ and $t_c''$ and an expression for $\gamma_c$ involving only elementary functions of $t_c$ (and no integrals).

In Section~\ref{sec4}, we shall show that for all $\tau\geq4$ and $q>2$
\beq \label{e20}
t_c<2\,{\rm ln}(q-1)~,~~~~~~t_c'<\tfrac32\,{\rm ln}(q-1)~,~~~~~~~t_c'' <{\rm ln}(q-1)~,
\eq
where we recall that
\beq \label{e21}
t_c''=t_{\ast}<t_c'=t_b<t_c~,
\eq
see \cite{ref1}, Figure~1. The proof of (\ref{e20}) uses the criteria in (\ref{e12}), (\ref{e13}) and (\ref{e14}), the representations (\ref{e16}), (\ref{e17}) and (\ref{e18}) of $\cak$, $\cak'$ and $\cak''$ and a convexity property of the function $(e^{tw}+q-1)^{-1}$, see (\ref{e19}), as a function of $w\geq1$.

In Section~\ref{sec5}, we shall improve the bounds on $t_c$ and $t_c'$ as follows. We have for $\tau\geq4$ and $q>2$
\beq \label{e22}
t_c<2\,\frac{\tau-2}{\tau-1}\,{\rm ln}(q-1)~.
\eq
The proof of this uses the observation that
\beq \label{e23}
\frac{\cak(2\,{\rm ln}(q-1))}{\cak'(2\,{\rm ln}(q-1))}=\frac{2\,{\rm ln}(q-1)}{\tau-1}~,
\eq
so that by $t_c<2\,{\rm ln}(q-1)$ and convexity of $\cak(t)$ in $t\geq t_c$
\beq \label{e24}
t_c<2\,{\rm ln}(q-1)-\frac{\cak(2\,{\rm ln}(q-1))}{\cak'(2\,{\rm ln}(q-1))}~,
\eq
the right-hand side of (\ref{e24}) being the first Newton iterand $t^{(1)}$ to compute $t_c$ starting with $t^{(0)}=2\,{\rm ln}(q-1)$. From (\ref{e23}), (\ref{e24}) we get (\ref{e22}).

To improve the bound on $t_c'$, we let $T$ be the unique positive solution $t$ of the equation
\beq \label{e25}
\frac{t\,e^t}{(e^t+q-1)^2}+\frac{1}{e^t+q-1}-\frac1q=0~.
\eq
Then it appears that
\beq \label{e26}
t_c'<T<\tfrac32\,{\rm ln}(q-1)~.
\eq
Furthermore, the choice of $T$ is such that
\beq \label{e27}
\frac{\cak'(T)}{\cak''(T)}=\frac{T}{\tau-2}~.
\eq
Hence, we have
\beq \label{e28}
T-\frac{\cak'(T)}{\cak''(T)}=\frac{\tau-3}{\tau-2}\,T~,
\eq
where the left-hand side of (\ref{e28}) is the first Newton iterand $t^{(1)}$ to compute $t_c'$ starting with $t^{(0)}=T$. While $\cak(t)$ is convex in $t\geq t_c$, convexity of $\cak'(t)$ in $t\in[t_c',T]$ does not always hold, and there are indeed cases that $\frac{\tau-3}{\tau-2}\,T<t_c'$, i.e., the right-hand side of (\ref{e28}) is not necessarily an upper bound of $t_c'$.

In Section~\ref{sec6}, we show the following counterparts to the inequalities (\ref{e20}) and (\ref{e22}). We have for any $\tau\geq4$
\beq \label{e29}
\lim_{q\pr\infty}\,\frac{t_c}{{\rm ln}(q-1)}=2\,\frac{\tau-2}{\tau-1}\,,~~ \lim_{q\pr\infty}\,\frac{t_c'}{{\rm ln}(q-1)}=1\,,~~ \lim_{q\pr\infty}\,\frac{t_c''}{{\rm ln}(q-1)}=1~.
\eq
Furthermore, it is shown that $t_c'<{\rm ln}(q-1)$ when $\tau=4$, $q>2$, and that $t_c'>{\rm ln}(q-1)$ when $\tau>4$ and $q$ is sufficiently large. We also discuss in Section~\ref{sec6} the conjecture that for $\tau\geq4$ and $q>2$
\beq \label{e30}
t_c>2\,\frac{\tau-5}{\tau-4}\,{\rm ln}(q-1)~.
\eq
By the first limit relation in (\ref{e29}), we have that (\ref{e30}) holds for $\tau\geq4$ and $q$ sufficiently large. We also give in Section~\ref{sec6} a consequence of the first limit relation in (\ref{e29}) for $\gamma_c=\exp(\beta_c)-1$, with $\beta_c$ the critical inverse temperature.

In Section~\ref{sec7}, we consider the limiting behaviour of $t_c''$ as $q\downarrow 2$. This is based on the observation that $t_c''=t_{\ast}$ is the unique positive zero of $\caf_0''$, so that by \cite{ref1}, Section~7.2, $t_c''$ is the unique positive zero $t$ of
\beq \label{e31}
\Phi(t):=\il_t^{\infty}\,x^{-\tau+3}\,a(x)\,dx~;~~~~~~a(x)=\frac{(q-1)\,e^x-e^{2x}} {(e^x+q-1)^3}\,,~~x\geq0~.
\eq
The function $a(x)$ is positive for $x\in[0,{\rm ln}(q-1))$ and negative for $x\in({\rm ln}(q-1),\infty)$, where ${\rm ln}(q-1)\downarrow0$ as $q\downarrow2$. Moreover, $\Phi(t)\pr\infty$ as $t\downarrow0$ in a $\tau$-dependent way. It thus appears that one has to distinguish between the cases
\beq \label{e32}
\mbox{(i)}~\tau=4\,,~~\mbox{(ii)}~4<\tau<5\,,~~\mbox{(iii)}~\tau=5\,,~~ \mbox{(iv)}~\tau>5
\eq
with respective decay behaviour of $t_c''$ as $b={\rm ln}(q-1)\downarrow0$
\beq \label{e33}
\mbox{(i)}~b\,\exp({-}K_1/b)\,,~~\mbox{(ii)}~K_2\,b^{\frac{1}{\tau-4}}\,,~~ \mbox{(iii)}~K_3\,b/|{\rm ln}\,b|\,,~~ \mbox{(iv)}~K_4\,b~,
\eq
where $K_1$, $K_2$, $K_3$, $K_4$ depend on $\tau$, and, interestingly, $K_4=\frac{\tau-5}{\tau-4}$.

Finally, in Section~\ref{sec8}, we consider the homogeneous case that arises when $\tau\pr\infty$, so that $f_W(w)\pr\delta(w-1)$. Denote by $t_{H,c}$, $t_{H,c}'$ and $t_{H,c}''$ the unique positive zeros of $\cak_H$, $\cak_H'$ and $\cak_H''$, respectively, with $\cak_H$ the $\cak$-function corresponding to this homogeneous case. As is well known, we have $t_{H,c}=2\,{\rm ln}(q-1)$. Now it turns out that $t_{H,c}'=T$, with $T$ the unique positive solution $t$ of (\ref{e25}). Furthermore, we have $t_{H,c}''={\rm ln}(q-1)$. Hence, the inequalities $t_c\leq2\,{\rm ln}(q-1)$, $t_c'\leq T$ and $t_c''<{\rm ln}(q-1)$, holding for the finite-$\tau$ Pareto case, are sharp in the sense that they hold with equality in the limit case $\tau\pr\infty$. While $t_{H,c}$ and $t_{H,c}''$ have simple closed forms, as just noted, such a thing does not hold for $t_{H,c}'$, and so a further analysis is required for the latter case. It is shown in Section~\ref{sec9} that $t_{H,c}'=T(q)$ increases in $q>2$ from 0 to $\infty$, and that $T(q)/{\rm ln}(q-1)$ decreases in $q>2$ from $3/2$ to 1 with $T(q)/{\rm ln}(q-1)\sim 1+{\rm ln}({\rm ln}(q-1))/{\rm ln}(q-1)$ as $q\pr\infty$.

\section{Proof of the representations (\ref{e16}), (\ref{e17}) and (\ref{e18})} \label{sec3}
\mbox{} \\[-9mm]

The proof of the representation (\ref{e16}) of $\cak$ has been given in \cite{ref1}, Appendix~C by using the definition (\ref{e6}) of $\cak$. Thus, by partial integration,
\begin{eqnarray} \label{e34}
& \mbox{} & \hspace*{-6mm}\dE\,[{\rm ln}(e^{tW}+q-1)]=\il_1^{\infty}\,(\tau-1)\,w^{-\tau}\,{\rm ln}(e^{tw}+q-1)\,dw \nonumber \\[3.5mm]
& & \hspace*{-6mm}=~\il_1^{\infty}\,{\rm ln}(e^{tw}+q-1)\,d({-}w^{-\tau+1})
={\rm ln}(e^t+q-1)+t\,\il_1^{\infty}\, \frac{w^{-\tau+1}\,e^{tw}} {e^{tw}+q-1}\,dw \nonumber \\[3.5mm]
& & \hspace*{-6mm}=~{\rm ln}(e^t+q-1)+\frac{t}{\tau-2}-(q-1)\,t\,\il_1^{\infty}\, \frac{w^{-\tau+1}\,dw}{e^{tw}+q-1}~.
\end{eqnarray}
Furthermore,
\beq \label{e35}
\caf_0(t)=1-\frac{q}{\dE\,[W]}\,\dE\,\Bigl[\frac{W}{e^{tW}+q-1}\Bigr]= 1-q(\tau-2)\,\il_1^{\infty}\,\frac{w^{-\tau+1}\,dw}{e^{tw}+q-1}~,
\eq
and the remaining integrals in (\ref{e34}) and (\ref{e35}) equal the $D$ in (\ref{e19}).

For proving the representation (\ref{e17}) for $\cak'$, one differentiates the expression (\ref{e16}) for $\cak$ with respect to $t$, and then one is left with the $D$-integral itself and its derivative with respect to $t$. As to the latter, we have
\begin{eqnarray} \label{e36}
& \mbox{} & \hspace*{-1cm}\frac{d}{dt}\,\left[\il_1^{\infty}\, \frac{w^{-\tau+1}\,dw}{e^{tw}+q-1} \right]=\frac{d}{dt}\,\left[t^{\tau-2}\,\il_t^{\infty}\,\frac{v^{-\tau+1}\,dv} {e^v+q-1} \right] \nonumber \\[3.5mm]
& & \hspace*{-1cm}=~(\tau-2)\,t^{\tau-3}\,\il_t^{\infty}\, \frac{v^{-\tau+1}\,dv}{e^v+q-1} -t^{\tau-2}\,\frac{t^{-\tau+1}}{e^t+q-1} \nonumber \\[3.5mm]
& & \hspace*{-1cm}=~\frac{\tau-2}{t}\,\il_1^{\infty}\,\frac{w^{-\tau+1}\,dw}{e^{tw}+q-1}- \frac{1}{t(e^t+q-1)}=\frac{\tau-2}{t}\,D-\frac{1}{t(e^t+q-1)}\,.
\end{eqnarray}

Finally, for proving the representation (\ref{e18}) of $\cak''$, one differentiates the expression (\ref{e17}) for $\cak'$ and uses (\ref{e36}) once more. 

The fact that in the representation of $\cak$, $\cak'$ and $\cak''$ only one and the same integral $D$ is required, is quite convenient, both from an analytic, see the following sections, and a numerical point of view. As to the latter point, both $t_c$ and $t_c'$ can be computed by a Newton iteration
\beq \label{e37}
t^{(j+1)}=t^{(j)}-\frac{\cael(t^{(j)})}{\cael'(t^{(j)})}\,,~~ j=0,1,...~;~~~~~~\cael=\cak~{\rm or}~\cak'~,
\eq
where an appropriate starting value $t^{(0)}$ should be chosen. For $\cael=\cak$, any upper bound of $t_c$ can be taken since $\cak(t)$ is convex in $t\in(t_c'',\infty)\supset [t_c,\infty)$. The computation of $t_c$ by Newton iteration has also been considered in \cite{ref1}, Appendix~C. For the Pareto case, there is given in \cite{ref1}, (C.10), a considerably more complicated expression for $\cak'$ than the one we presently have in (\ref{e17}). Moreover, the starting value $t^{(0)}$ that is used in \cite{ref1}, Appendix~C, is $t^{(0)}= \frac{\tau-2}{\tau-1}~\frac{2q\,{\rm ln}\,q}{q-1}$ which exceeds $2\,\frac{\tau-2}{\tau-1}\,{\rm ln}(q-1)$, see (\ref{e22}).

We can compute, in principle, also $t_c'$ by Newton iteration, but there is here the complication that $\cak'$ is not convex on the whole range $[t_c',\infty)$. For it is seen from the expression for $\cak''(t)$ in (\ref{e18}) that $\cak''(t)\pr0$ as $t\pr\infty$, and so $\cak'''(t)<0$ for certain large $t$. With $T$ the upper bound of $t_c'$ obtained in (\ref{e25}), (\ref{e26}), it appears (at least numerically) that $\cak'(t)$ is convex in $t\in[t_c',T]$ when $q$ is not very large compared to $\tau$.
\newpage
\section{Proof of the simple bounds in (\ref{e20})} \label{sec4}
\mbox{} \\[-9mm]

In this section we show that
\beq \label{e38}
t_c<2\,{\rm ln}(q-1)~,~~~~~~t_c'<\tfrac32\,{\rm ln}(q-1)~,~~~~~~ t_c''<{\rm ln}(q-1)
\eq
for all $\tau\geq4$ and $q>2$. According to the criteria in (\ref{e12}), (\ref{e13}), (\ref{e14}), we have that (\ref{e38}) is equivalent with
\beq \label{e39}
\cak(2\,{\rm ln}(q-1))>0~,~~~~~~ \cak'(\tfrac32\,{\rm ln}(q-1))>0~,~~~~~~ \cak''({\rm ln}(q-1))>0~.
\eq
To establish (\ref{e39}), we consider for $\alpha\geq1$ and $q>2$, $t=\alpha\,{\rm ln}(q-1)$ the expression
\beq \label{e40}
(q-1)\,D=(q-1)\,\il_1^{\infty}\,\frac{w^{-\tau+1}\,dw}{e^{tw}+q-1}=\il_1^{\infty}\, \frac{w^{-\tau+1}\,dw}{(q-1)^{\alpha w-1}+1}~,
\eq
where $D=D(t)$ with $t=\alpha\,{\rm ln}(q-1)$ in (\ref{e19}). The function
\beq \label{e41}
f(w):=\frac{1}{(q-1)^{\alpha w-1}+1}=\frac{1}{\exp((\alpha w-1)\,{\rm ln}(q-1))+1}
\eq
is positive, decreasing and strictly convex in $w>1$. Indeed, we have $\alpha\geq1$, ${\rm ln}(q-1)>0$ and so $z:=(\alpha w-1)\,{\rm ln}(q-1)>0$ for $w>1$, while
\beq \label{e42}
\Bigl(\frac{1}{e^z+1}\Bigr)''=\Bigl(\frac{-e^z}{(e^z+1)^2}\Bigr)'= e^z\,\frac{e^z-1}{(e^z+1)^3}>0~.
\eq
It follows that
\beq \label{e43}
f(w)>f(1)+f'(1)(w-1)~,~~~~~~w>1~,
\eq
and so, by straightforward integration,
\beq \label{e44}
(q-1)\,D>\il_1^{\infty}\,w^{-\tau+1}(f(1)+f'(1)(w-1))\,dw=\frac{f(1)}{\tau-2}+\frac{f'(1)} {(\tau-2)(\tau-3)}~.
\eq

Now we show the first inequality in (\ref{e39}) for which we make the choice $\alpha=2$. We have from (\ref{e16})
\begin{eqnarray} \label{e45}
& \mbox{} & \cak(2\,{\rm ln}(q-1)) \nonumber \\[3mm]
& & =~\frac{\tau-2}{\tau-1}\,{\rm ln}\Bigl( \frac{(q-1)^2+q-1}{q}\Bigr)+ \Bigl(\frac{1}{\tau-1}-\frac{q+1}{2q}\Bigr)\,2\,{\rm ln}(q-1) \nonumber \\[3mm]
& & \hspace*{6mm}+~\frac{(\tau-2)(\tau-3)}{2(\tau-1)}\,2\,{\rm ln}(q-1)(q-1)\,D \nonumber \\[3mm]
& & =~{\rm ln}(q-1)\Bigl(\frac{\tau-2}{\tau-1}+2\Bigl(\frac{1}{\tau-1}- \frac{q+1}{2q}\Bigr) +\frac{(\tau-2)(\tau-3)}{\tau-1}\,(q-1)\,D\Bigr) \nonumber \\[3mm]
& & =~{\rm ln}(q-1)\Bigl(\frac{1}{\tau-1}-\frac1q+\frac{(\tau-2)(\tau-3)}{\tau-1}\,(q-1)\,D\Bigr)~.
\end{eqnarray}
With $\alpha=2$ in (\ref{e41}), we have $f(w)=((q-1)^{2w-1}+1)^{-1}$, and so 
\beq \label{e46}
f(1)=\frac1q~,~~~~~~f'(1)=\frac{-2(q-1)\,{\rm ln}(q-1)}{q^2}~.
\eq
Hence, from (\ref{e44}),
\begin{eqnarray} \label{e47}
\frac{(\tau-2)(\tau-3)}{\tau-1}\,(q-1)\,D & > &  \frac{\tau-3}{\tau-1}\,f(1)+\frac{1}{\tau-1}\,f'(1) \nonumber \\[3mm]
& = & \frac{\tau-3}{\tau-1}~\frac1q-\frac{2}{\tau-1}~\frac{(q-1)\,{\rm ln}(q-1)}{q^2}~.
\end{eqnarray}
Then from (\ref{e45})
\begin{eqnarray} \label{e48}
& \mbox{} & \cak(2\,{\rm ln}(q-1)) \nonumber \\[3mm]
& & >~{\rm ln}(q-1)\Bigl(\frac{1}{\tau-1}-\frac1q+\frac{\tau-3}{\tau-1}~\frac1q-\frac{2}{\tau-1}~ \frac{(q-1)\,{\rm ln}(q-1)}{q^2}\Bigr) \nonumber \\[3mm]
& & =~\frac{{\rm ln}(q-1)}{\tau-1}\,\Bigl(1-\frac2q-\frac{2(q-1)\,{\rm ln}(q-1)}{q^2}\Bigr)
\nonumber \\[3mm]
& & =~\frac{{\rm ln}(q-1)}{(\tau-1)\,q^2}\,((q-1)^2-1-2(q-1)\,{\rm ln}(q-1))>0
\end{eqnarray}
since $x^2-1-2x\,{\rm ln}\,x>0$ for $x=q-1>1$.

We next show the second inequality in (\ref{e39}), for which we make the choice $\alpha=3/2$. We have from (\ref{e17}) with $t=\frac32\,{\rm ln}(q-1)$, $\exp(t)=(q-1)^{3/2}$
\beq \label{e49}
\cak'(\tfrac32\,{\rm ln}(q-1))=\frac{q-1}{2q}-\tfrac12\:\frac{\tau-2}{(q-1)^{1/2}+1}+ \tfrac12\,(\tau-2)(\tau-3)(q-1)\,D~.
\eq
According to (\ref{e44})
\beq \label{e50}
\tfrac12\,(\tau-2)(\tau-3)(q-1)\,D> \tfrac12\,(\tau-3)\,f(1)+\tfrac12\,f'(1)~,
\eq
where now $f(w)=((q-1)^{\frac32 w-1}+1)^{-1}$. Then from
\beq \label{e51}
f(1)=\frac{1}{(q-1)^{1/2}+1}~,~~~~~~f'(1)=\frac{-\,\tfrac32\,(q-1)^{1/2}\,{\rm ln}(q-1)} {((q-1)^{1/2}+1)^2}~,
\eq
we see that
\beq \label{e52}
\cak'(\tfrac32\,{\rm ln}(q-1))>\tfrac12\,\Bigl(\frac{q-1}{q}-\frac{1}{(q-1)^{1/2}+1}- \frac{\tfrac32\,(q-1)^{1/2}\,{\rm ln}(q-1)}{((q-1)^{1/2}+1)^2}\Bigr)~.
\eq
Hence, we have $\cak'(\frac32\,{\rm ln}(q-1))>0$ when we can show that
\beq \label{e53}
\frac{x}{1+x}-\frac{1}{1+x^{1/2}}- \frac{\frac32\,x^{1/2}\,{\rm ln}\,x}{(1+x^{1/2})^2}>0 ~,~~~~~~x=q-1>1~.
\eq
Setting $y=x^{1/2}>1$, we consider
\beq \label{e54}
\varp(y):=y^2(1+y)^2-(1+y)(1+y^2)-3y(1+y^2)\,{\rm ln}\,y~,~~~~~~y\geq1~.
\eq
Then (\ref{e53}) is equivalent with $\varp(y)>0$, $y>1$. We have
\beq \label{e55}
\varp(1)=\varp'(1)=\varp''(1)=\varp'''(1)=0~;~~~~~~\varp''''(y)=24-\frac{18}{y}-\frac{6}{y^3}>0\,,~~ y>1~,
\eq
and so $\varp(y)>0$, $y>1$ indeed holds.

We finally show the third inequality in (\ref{e39}), for which we make the choice $\alpha=1$. We have from (\ref{e18}) with $t={\rm ln}(q-1)$, $\exp(t)=q-1$
\begin{eqnarray} \label{e56}
& \mbox{} & \hspace*{-4mm}\cak''({\rm ln}(q-1)) \nonumber \\[3mm]
& & \hspace*{-4mm}=~\tfrac12\,(\tau-2)\Bigl(\frac{(q-1)^2}{(2(q-1))^2}- \frac{(\tau-3)(q-1)}{2(q-1)\,{\rm ln}(q-1)}+\frac{(\tau-2)(\tau-3)(q-1)\,D} {{\rm ln}(q-1)} \Bigr) \nonumber \\[3mm]
& & \hspace*{-4mm}=~\tfrac12\,(\tau-2)\Bigl(\tfrac14-\frac{\tau-3}{2\,{\rm ln}(q-1)}+ \frac{(\tau-2)(\tau-3)(q-1)\,D}{{\rm ln}(q-1)}\Bigr)~.
\end{eqnarray}
According to (\ref{e44})
\beq \label{e57}
(\tau-2)(\tau-3)(q-1)\,D>(\tau-3)\,f(1)+f'(1)~,
\eq
where now $f(w)=((q-1)^{w-1}+1)^{-1}$. Then from
\beq \label{e58}
f(1)=\tfrac12~,~~~~~~f'(1)=-\,\tfrac14\,{\rm ln}(q-1)~,
\eq
we see that
\beq \label{e59}
\cak''({\rm ln}(q-1))>\tfrac12\,(\tau-2)\Bigl(\tfrac14-\frac{\tau-3} {2\,{\rm ln}(q-1)}+ \frac{\tfrac12\,(\tau-3)-\tfrac14\,{\rm ln}(q-1)}{{\rm ln}(q-1)}\Bigr)=0~,
\eq
as required.

A quick proof of $t_c''<{\rm ln}(q-1)$ follows from the characterization of $t_c''$ as the unique positive zero of $\Phi(t)$ in (\ref{e31}) which is obviously negative for $t\geq{\rm ln}(q-1)$ since $a(x)<0$ for $x>{\rm ln}(q-1)$.

\section{Sharpening of the simple bounds for $t_c$ and $t_c'$} \label{sec5}
\mbox{} \\[-9mm]

We start by showing that
\beq \label{e60}
t_c<2\,\frac{\tau-2}{\tau-1}\,{\rm ln}(q-1)<2\,{\rm ln}(q-1)
\eq
for all $\tau\geq4$ and $q>2$. The second inequality is obvious. Since $\cak(t)$ is convex in $t\geq t_c>t_c''$, we have for any upper bound $T$ of $t_c$ that
\beq \label{e61}
t_c<T-\frac{\cak(T)}{\cak'(T)}<T~,
\eq
where the middle member is the first Newton iterand $t^{(1)}$ to compute $t_c$ starting with $t^{(0)}=T$. We take $T=2\,{\rm ln}(q-1)$, and we have from (\ref{e45})
\beq \label{e62}
\cak(2\,{\rm ln}(q-1))=\frac{{\rm ln}(q-1)}{\tau-1}\,\Bigl(1-\frac{\tau-1}{q}+(\tau-2)(\tau-3)(q-1)\,D\Bigr)~.
\eq
At the same time, we compute from (\ref{e17})
\begin{eqnarray} \label{e63}
\cak'(2\,{\rm ln}(q-1)) & = & \tfrac12\,(q-1)\Bigl(\frac1q-\frac{\tau-2} {(q-1)^2+q-1}+(\tau-2)(\tau-3)\,D\Bigr) \nonumber \\[3mm]
& = & \tfrac12\,\Bigl(1-\frac1q-\frac{\tau-2}{q}+(\tau-2)(\tau-3)(q-1)\,D\Bigr) \nonumber \\[3mm]
& = & \tfrac12\,\Bigl(1-\frac{\tau-1}{q}+(\tau-2)(\tau-3)(q-1)\,D\Bigr)~,
\end{eqnarray}
in which the two $D$'s in (\ref{e62}) and (\ref{e63}) are equal. Hence,
\beq \label{e64}
\cak(2\,{\rm ln}(q-1))=2\,\frac{{\rm ln}(q-1)}{\tau-1}\,\cak'(2\,{\rm ln}(q-1))~,
\eq
and so, by (\ref{e61}),
\beq \label{e65}
t_c<2\,{\rm ln}(q-1)-\frac{\cak(2\,{\rm ln}(q-1))}{\cak'(2\,{\rm ln}(q-1))}= 2\,\frac{\tau-2}{\tau-1}\,{\rm ln}(q-1)~.
\eq

We next consider sharpening the simple upper bound $\frac32\,{\rm ln}(q-1)$ of $t_c'$. Our initial aim is to find an upper bound $T$ of $t_c'$ such that $\cak'(T)/\cak''(T)$ takes a simple form, just as we succeeded in doing so to improve the simple upper bound $2\,{\rm ln}(q-1)$ of $t_c$. Such a simple form for $\cak'(T)/\cak''(T)$ arises from (\ref{e17}) and (\ref{e18}) when we manage to take $T=t>0$ such that
\beq \label{e66}
\frac1q-\frac{\tau-2}{e^t+q-1}=\frac{t\,e^t}{(e^t+q-1)^2}-\frac{\tau-3} {e^t+q-1}~,
\eq
the two $D$'s in (\ref{e17}) and (\ref{e18}) being the same. Thus, with this $T=t$, we get
\beq \label{e67}
\frac{\cak'(T)}{\cak''(T)}=\frac{T}{\tau-2}~.
\eq
The first Newton iterand $t^{(1)}$ to compute $t_c'$ starting with $t^{(0)}=T$ is given by
\beq \label{e68}
T-\frac{\cak'(T)}{\cak''(T)}=\frac{\tau-3}{\tau-2}\,T~.
\eq

We now detail this approach further. The equation in (\ref{e66}) is the same as
\beq \label{e69}
\frac{t\,e^t}{(e^t+q-1)^2}+\frac{1}{e^t+q-1}-\frac1q=0
\eq
from which $\tau$ has disappeared altogether. Multiplying through by $(e^t+q-1)^2$, we get from (\ref{e69}) the equation
\beq \label{e70}
\psi(e^t)=e^t(1+t)+q-1-\frac1q\,(e^t+q-1)^2=0
\eq
in which
\beq \label{e71}
\psi(y):=y(1+{\rm ln}\,y)+q-1-\frac1q\,(y+q-1)^2~,~~~~~~y\geq1~.
\eq
We shall show that $\psi(y)$ has a unique zero $Y>1$. We have, since $q>2$,
\beq \label{e72}
\psi(1)=\psi'(1)=0<\psi''(1)=1-\frac2q~,
\eq
and
\beq \label{e73}
\psi''(y)=\frac1y-\frac2q~,~~~~~~y\geq1~.
\eq
Therefore, $\psi''(y)>0$ for $y\in[1,\frac12\,q)$ and $\psi''(y)<0$ for $y\in(\frac12\,q,\infty)$. Thus, $\psi(y)$ is positive, increasing and convex in $y\in(1,\frac12\,q)$, and $\psi(y)$ is concave in $y\in(\frac12\,q,\infty)$, while we see from (\ref{e71}) that $\psi(y)\pr{-}\infty$ as $y\pr\infty$. We conclude that $\psi(y)$ has a unique positive zero $Y$ with $Y>\frac12\,q>1$.

We shall show in Appendix~A that
\beq \label{e74}
\psi(q-1)>0>\psi((q-1)^{3/2})~.
\eq
Hence
\beq \label{e75}
q-1<Y<(q-1)^{3/2}~.
\eq
Therefore, $T={\rm ln}\,Y$ is the unique positive solution $t$ of (\ref{e70}), and 
\beq \label{e76}
{\rm ln}(q-1)<T<\tfrac32\,{\rm ln}(q-1)~.
\eq
From the bound $t_c''<{\rm ln}(q-1)$, we have $t_c''<T$, and so, from (\ref{e14}),
\beq \label{e77}
\cak''(T)>0~.
\eq
The choice of $T$ is such that the two quantities in the brackets (~~) at the right-hand sides of the representations (\ref{e17}) and (\ref{e18}) of $\cak'$ and $\cak''$ are the same, see (\ref{e66}), for $t=T$. From $\cak''(T)>0$, we then conclude that $\cak'(T)>0$ as well, and so, by (\ref{e13}), $t_c'<T$. Therefore,
\beq \label{e78}
t_c'<T<\tfrac32\,{\rm ln}(q-1)~,
\eq
and so $T$ indeed sharpens $\frac32\,{\rm ln}(q-1)$ as an upper bound of $t_c'$. 

It is now tempting to ask whether $\frac{\tau-3}{\tau-2}\,T$ is an upper bound of $t_c'$ as well. This would be the case when $\cak'(t)$ were convex in $t\in[t_c',T]$, i.e., when $\cak''(t)$ were increasing in $t\in[t_c',T]$. However, this does not hold always, as we illustrate now.
\newpage
\begin{eqnarray}
q=20,~\tau=6 & : & t_c'=3.1829\,,~~T=4.1914\,,~~\frac{\tau-3}{\tau-2}\,T=3.1436< t_c'\,,~{\rm and} \nonumber \\[2mm]
& & \cak''(t)~\mbox{is decreasing in}~t\in[3.7,4.2] \nonumber \\[6mm]
q=20,~\tau=11 & : & t_c'=3.7205\,,~~T=4.194\,,~~\frac{\tau-3}{\tau-2}\,T=3.7257>t_c'\,,~{\rm and} \nonumber \\[2mm]
& & \cak''(t)~\mbox{is decreasing in}~t\in[4.09,4.20] \nonumber \\[6mm]
q=20,~\tau=18 & : & t_c'=3.9245\,,~~T=4.1914\,,~~\frac{\tau-3}{\tau-2}\,T=3.9294 >t_c'\,,~{\rm and} \nonumber \\[2mm]
& & \cak''(t)~\mbox{is increasing in}~t\in[t_c',4.25] \nonumber
\end{eqnarray}

\section{Limiting behaviour of $t_c$, $t_c'$ and $t_c''$ as $q\pr\infty$} \label{sec6}
\mbox{} \\[-9mm]

We shall now establish the limit relations in (\ref{e29}). To show the first one of these, viz.\ $t_c/{\rm ln}(q-1)\pr 2(\tau-2)/(\tau-1)$ as $q\pr\infty$, it is by (\ref{e12}) and (\ref{e22}) sufficient to show that for any $\alpha\in(1,2(\tau-2)/(\tau-1))$ we have
\beq \label{e79}
\cak(\alpha\,{\rm ln}(q-1))<0
\eq
when $q$ is sufficiently large. So let $\alpha\in(1,2(\tau-2)/(\tau-1))$. We use (\ref{e16}) with $t=\alpha\,{\rm ln}(q-1)$, $\exp(t)=(q-1)^{\alpha}$, and we write
\begin{eqnarray} \label{e80}
\frac{\tau-2}{\tau-1}\,{\rm ln}\Bigl(\frac{e^t+q-1}{q}\Bigr) & = &
\frac{\tau-2}{\tau-1}\,{\rm ln}\Bigl(\frac{(q-1)^{\alpha}+q-1}{q}\Bigr) \nonumber \\[3mm]
& = & \frac{\tau-2}{\tau-1}\,{\rm ln}\Bigl[(q-1)^{\alpha-1}\cdot\frac{q-1}{q}\cdot (1+(q-1)^{1-\alpha})\Bigr] \nonumber \\[3mm]
& = & \frac{\tau-2}{\tau-1}\,(\alpha-1)\,{\rm ln}(q-1)+O\Bigl( \frac1q+\Bigl(\frac1q\Bigr)^{\alpha-1}\Bigr)~. \nonumber \\
& & \mbox{}
\end{eqnarray}
Furthermore,
\begin{eqnarray} \label{e81}
& \mbox{} & \hspace*{-1.2cm}\frac{(\tau-2)(\tau-3)}{2(\tau-1)}\,t(q-1)\,D \nonumber \\[3mm]
& & \hspace*{-1.2cm}=~\frac{(\tau-2)(\tau-3)}{2(\tau-1)}\,\alpha\,{\rm ln}(q-1)\,\il_1^{\infty}\, \frac{w^{-\tau+1}\,dw}{(q-1)^{\alpha w-1}+1}=O\Bigl(\frac{{\rm ln}(q-1)} {(q-1)^{\alpha-1}} \Bigr)~.
\end{eqnarray}
Therefore,
\begin{eqnarray} \label{e82}
& \mbox{} & \hspace*{-8mm}\cak(\alpha\,{\rm ln}(q-1)) \nonumber \\[3mm]
& & \hspace*{-8mm}=~\frac{\tau-2}{\tau-1}\,(\alpha-1)\,{\rm ln}(q-1)+\Bigl(\frac{1}{\tau-1}- \frac{q+1}{2q}\Bigr)\,\alpha\,{\rm ln}(q-1)+O\Bigl(\frac{{\rm ln}(q-1)}{(q-1)^{\alpha-1}} \Bigr) \nonumber \\[3mm]
& & \hspace*{-8mm}=~\tfrac12\Bigl(\alpha-2\,\frac{\tau-2}{\tau-1}-\frac{\alpha}{q}+O\Bigl( \frac{1}{(q-1)^{\alpha-1}}\Bigr)\Bigr)\,{\rm ln}(q-1)~.
\end{eqnarray}
Since $1<\alpha<2(\tau-2)/(\tau-1)$, we see that $\cak(\alpha\,{\rm ln}(q-1))<0$ indeed, when $q$ is sufficiently large.

We conjecture the lower bound for $\tau\geq5$ and all $q>2$
\beq \label{e83}
t_c>2\,\frac{\tau-5}{\tau-4}\,{\rm ln}(q-1)~,
\eq
and this seems to be a very sharp lower bound when $\tau$ is not too small and $q$ is not too large. From $t_c/{\rm ln}(q-1)\pr 2(\tau-2)/(\tau-1)$ as $q\pr\infty$ and $(\tau-2)/(\tau-1)>(\tau-5)/(\tau-4)$, we see that (\ref{e83}) holds when $\tau\geq4$ and $q$ is sufficiently large. For $\tau=6$, we would have $t_c>{\rm ln}(q-1)$.

In fact, a limited numerical exploration with some other pdfs (like the exponential $\exp({-}w)\,\cax_{[0,\infty)}(w)$ and $\cax_{[0,1)}(w)$) shows that there may hold more generally
\beq \label{e84}
2\,\frac{\mu_3}{\mu_4}\,{\rm ln}(q-1)<t_c<2\,\frac{\mu_0}{\mu_1}\,{\rm ln}(q-1)~, ~~~~~~q>2~,
\eq
where $\mu_n$ is the $n^{{\rm th}}$ moment,
\beq \label{e85}
\mu_n=\il_0^{\infty}\,w^n\,f_W(w)\,dw~,~~~~~~n=0,1,...
\eq
of the pdf $f_W$. Note that by a continuous version of Lehmer's inequality, we have $\mu_{p-1}/\mu_p>\mu_{q-1}/\mu_q$ when $p<q$ (see \cite{ref4}), whence $\mu_3/\mu_4<\mu_0/\mu_1$.

In Appendix~B, we shall show that
\beq \label{e85+1}
t_c=2\,\frac{\tau-2}{\tau-1}\,(1+O(1/p^{\frac{\tau-3}{\tau-1}}))\,\rln\,p~,~~~~~~ p\pr\infty~,
\eq
where $p=q-1$. Hence, $0<2(\tau-2)/(\tau-1)\cdot\rln(q-1)-t_c\pr0$ exponentially fast in $\rln(q-1)$ as $q\pr\infty$.

We now give a consequence of the limit relation satisfied by $t_c$ for $\gamma_c=\exp(\beta_c)-1$ with $\beta_c$ the critical inverse temperature. We have
\beq \label{e86}
\gamma_c=\frac{t_c}{\caf_0(t_c)}~,
\eq
where $\caf_0$ is given by (\ref{e1}). In the Pareto case, we can elaborate $\caf_0(t)$ as
\beq \label{e87}
\caf_0(t)=1-q(\tau-2)\,\il_1^{\infty}\,\frac{w^{-\tau+1}\,dw}{e^{tw}+q-1}~,~~~~~~ t\geq0~.
\eq
Therefore,
\beq \label{e88}
1-\frac{q}{e^t+q-1}<\caf_0(t)<1~,~~~~~~t\geq0~,
\eq
and so
\beq \label{e89}
t_c<\gamma_c<t_c\Bigl(1+\frac{q}{e^{t_c}-1}\Bigr)~.
\eq
With $\alpha\in(1,2(\tau-2)/(\tau-1))$ fixed, we have
\beq \label{e90}
\alpha\,{\rm ln}(q-1)<t_c<2\,\frac{\tau-2}{\tau-1}\,{\rm ln}(q-1)
\eq
when $q$ is sufficiently large, and so
\beq \label{e91}
\gamma_c<t_c\Bigl(1+\frac{q}{(q-1)^{\alpha}-1}\Bigr)~,~~~~~~q\pr\infty~.
\eq

We shall next express $\gamma_c$, given by (\ref{e86}), as an elementary function of $x_c$, where we write $t_c=x_c\,\rln\,p$ with $p=q-1$. The formula (\ref{e16}) for $\cak(t)$ and (\ref{e35}) for $\caf_0(t)$ both contain the integral $D(t)$ of (\ref{e19}). Since $\cak(t_c)=0$, we see that $D(t_c)$, and hence $\caf_0(t_c)$, assume the form of an elementary function of $t_c$. The details are as follows. With $t=x\,\rln\,p$, we have
\beq \label{e91+1}
\rln\Bigl(\frac{e^t+q-1}{q}\Bigr)=\rln\Bigl(\frac{p^x+p}{p+1}\Bigr)=(x-1)\,\rln\,p+r(x)~,
\eq
where
\beq \label{e91+2}
r(x)=\rln\Bigl(\frac{1+1/p^{x-1}}{1+1/p}\Bigr)~,~~~~~~x>0~.
\eq
Then, from $\cak(t_c)=0$, we get
\begin{eqnarray} \label{e91+3}
0 & = & \frac{\tau-2}{\tau-1}\,((x-1)\,\rln\,p+r(x))+\Bigl(\frac{1}{\tau-1}- \frac{p+2}{2(p+1)}\Bigr)\,x\,\rln\,p \nonumber \\[3mm]
& & +~\frac12~\frac{\tau-3}{\tau-1}\cdot x\,\rln\,p\cdot(\tau-2)\,p\,D(t)~,~~~~~~ x=x_c\,,~~t=t_c~.
\end{eqnarray}
Hence,
\begin{eqnarray} \label{e91+4}
(\tau-2)\,p\,D(t) & = & -\,\frac{\dfrac{\tau-2}{\tau-1}\,((x-1)\,\rln\,p+r(x))+ \Bigl(\dfrac{1}{\tau-1}-\dfrac{p+2}{2(p+1)}\Bigr)\,x\,\rln\,p} {\dfrac12\dfrac{\tau-3}{\tau-1}\,x\,\rln\,p} \nonumber \\[3mm]
& = & -\,\frac{\tau-1}{\tau-3}~\frac{p}{p+1}+2\,\frac{\tau-2}{\tau-3}~\frac1x\, \Bigl(1-\frac{r(x)}{\rln\,p}\Bigr)~, \nonumber \\[3mm]
& & \hspace*{6cm}x=x_c\,,~~t=t_c~.
\end{eqnarray}
Therefore,
\begin{eqnarray} \label{e91+5}
\caf_0(t) & = & 1-\frac{p+1}{p}\,(\tau-2)\,p\,D(t)=2\,\frac{\tau-2}{\tau-3}\, \Bigl(1-\frac{p+1}{p}~\frac1x\,\Bigl(1-\frac{r(x)}{\rln\,p}\Bigr)\Bigr)~, \nonumber \\[3mm]
& & \hspace*{7.5cm}x=x_x\,,~~t=t_c~,
\end{eqnarray}
It follows then from (\ref{e86}) that
\beq \label{e91+6}
\gamma_c=\frac12~\frac{\tau-3}{\tau-2}~\frac{x}{x-\dfrac{p+1}{p}\,\Bigl(1-\dfrac{r(x)} {\rln\,p}\Bigr)}\,t_c~,~~~~~~x=x_c~.
\eq

Bounds, approximations, asymptotics and availability of computation schemes for $x_c$ and $t_c$ have a direct impact on $\gamma_c$ via (\ref{e91+6}). For instance, by (\ref{e85+1}) and (\ref{e91+2}), with $x=x_c$, we have
\begin{eqnarray} \label{e91+7}
\frac{x_c}{x_c-\dfrac{p+1}{p}\,\Bigl(1-\dfrac{r(x)}{\rln\,p}\Bigr)} & = & \frac{2(\tau-2)/(\tau-1)}{2(\tau-2)/(\tau-1)-1}\,(1+O(1/p^{\frac{\tau-3}{\tau-1}})) \nonumber \\[3mm]
& = & 2\,\frac{\tau-2}{\tau-3}\,(1+O(1/p^{\frac{\tau-3}{\tau-1}}))\,,~~~p\pr\infty~.
\end{eqnarray}
Hence, by (\ref{e91+6})
\beq \label{e91+8}
\gamma_c=t_c(1+O(1/p^{\frac{\tau-3}{\tau-1}}))~,
\eq
compare (\ref{e91}).

We next consider the limiting behaviour of $t_c'$ when $\tau\geq4$ and $q\pr\infty$. We first show that for any $\alpha>1$
\beq \label{e92}
\cak'(\alpha\,{\rm ln}(q-1))>0
\eq
when $q$ is sufficiently large. Hence, by (\ref{e13}), we have for any $\alpha>1$ that $t_c'<\alpha\,{\rm ln}(q-1)$ when $q$ is sufficiently large. So let $\alpha>1$. We have from (\ref{e17}), (\ref{e19})
\beq \label{e93}
\cak'(\alpha\,{\rm ln}(q-1))>\frac{q-1}{2q}-\frac{\tfrac12\,(\tau-2)} {(q-1)^{\alpha-1}+1} \pr\tfrac12~,~~~~~~q\pr\infty~,
\eq
since $\alpha>1$ and $D$ in (\ref{e19}) is positive. Hence, (\ref{e92}) holds when $\alpha>1$ and $q$ is sufficiently large.

Now let $\alpha\in(0,1)$. We shall show that
\beq \label{e94}
t_c'>\alpha\,{\rm ln}(q-1)
\eq
when $q$ is sufficiently large. For this, we use the result of the third limit relation in (\ref{e29}), to be shown below, so that $t_c''>\alpha\,{\rm ln}(q-1)$ when $q$ is sufficiently large. Since $t_c'>t_c''$, we thus get (\ref{e94}) when $\alpha\in(0,1)$ and $q$ is sufficiently large. From this and the above, it is concluded that $t_c'/{\rm ln}(q-1)\pr1$ as $q\pr\infty$ for any $\tau\geq4$, and this is the second limit relation in (\ref{e29}).

We shall now consider $\cak'({\rm ln}(q-1))$, and we shall show that 
\beq \label{e95}
t_c'>{\rm ln}(q-1)~,~~~~~~\tau>4~,
\eq
when $q$ is sufficiently large, and that
\beq \label{e96}
t_c'<{\rm ln}(q-1)~,~~~~~~\tau=4~,
\eq
for all $q>2$. We compute from (\ref{e17}) and (\ref{e19}) for $\tau\geq4$ and $q>2$
\begin{eqnarray} \label{e97}
& \mbox{} & \cak'({\rm ln}(q-1)) \nonumber \\[3mm]
& & =~\tfrac12\,(q-1)\left(\frac1q-\frac{\tau-2}{2(q-1)} +(\tau-2)(\tau-3)\,\il_1^{\infty}\, \frac{w^{-\tau+1}\,dw}{(q-1)^w+q-1}\right) \nonumber \\[3mm]
& & =~\tfrac12\,\left(2-\frac1q-\tfrac12\,\tau+(\tau-2)(\tau-3)\,\il_1^{\infty}\, \frac{w^{-\tau+1}\,dw}{(q-1)^{w-1}+1}\right)~.
\end{eqnarray}
We have by dominated convergence
\beq \label{e98}
\lim_{q\pr\infty}\,\il_1^{\infty}\,\frac{w^{-\tau+1}\,dw}{(q-1)^{w-1}+1}=0~.
\eq
Hence, when $\tau>4$, we have
\beq \label{e99}
\lim_{q\pr\infty}\,\cak'({\rm ln}(q-1))=\tfrac12\,(2-\tfrac12\,\tau)<0~,
\eq
and so (\ref{e95}) holds when $q$ is sufficiently large.

When $\tau=4$, the limit in (\ref{e99}) equals 0, and we have to proceed more carefully. For $\tau=4$, we have from (\ref{e97})
\beq \label{e100}
2\,\cak'({\rm ln}(q-1))={-}\,\frac1q+2\,\il_1^{\infty}\, \frac{w^{-3}\,dw}{(q-1)^{w-1}+1}~.
\eq
With $t={\rm ln}(q-1)$, $q=\exp(t)+1$, we have
\beq \label{e101}
2q\,\il_1^{\infty}\,\frac{w^{-3}\,dw}{1+(q-1)^{w-1}}=\il_0^{\infty}\, \frac{e^t+1}{e^{ty}+1} \,2(1+y)^{-3}\,dy~,
\eq
and we shall show that this exceeds 1 for all $q>2$, so that $\cak'({\rm ln}(q-1))>0$ and thus $t_c'<{\rm ln}(q-1)$ for all $q>2$ and $\tau=4$. The function $g(y):=(e^t+1)/(e^{ty}+1)$, $y\geq0$, is for any $t>0$ positive, decreasing and convex in $y\geq0$, and we have
\beq \label{e102}
g(1)=1~,~~~~~~g'(1)=\frac{-t\,e^t}{e^t+1}~.
\eq
Hence,
\beq \label{e103}
\frac{e^t+1}{e^{ty}+1}>1-\frac{t\,e^t}{e^t+1}\,(y-1)~,~~~~~~y\geq0\,,~~y\neq 1~.
\eq
We then have
\begin{eqnarray} \label{e104}
& \mbox{} & \il_0^{\infty}\,\frac{e^t+1}{e^{ty}+1}\,2(1+y)^{-3}\,dy> \il_0^{\infty}\, \Bigl(1-\frac{t\,e^t}{e^t+1}\,(y-1)\Bigr)\,2(1+y)^{-3}\,dy \nonumber \\[3.5mm]
& & =~\il_0^{\infty}\,\Bigl(1-\frac{t\,e^t}{e^t+1}\,(y-1)\Bigr)\,d\Bigl(\frac{-1}{(1+y)^2}\Bigr) \nonumber \\[3.5mm]
& & =~\Bigl(1-\frac{t\,e^t}{e^t+1}\,(y-1)\Bigr)\,\left. \frac{-1}{(1+y)^2}\right|_0^{\infty} +\il_0^{\infty}\,\frac{-t\,e^t}{e^t+1}~\frac{dy}{(1+y)^2} \nonumber \\[3.5mm]
& & =~1+\frac{t\,e^t}{e^t+1}-\frac{t\,e^t}{e^t+1}=1~.
\end{eqnarray}

We finally show the third limit relation in (\ref{e29}). Since $t_c''<{\rm ln}(q-1)$, it is sufficient to let $\tau\geq4$ and $\alpha\in(0,1)$, and to show that $\cak''(\alpha\,{\rm ln}(q-1))<0$ when $q$ is sufficiently large, see (\ref{e14}). With $t=\alpha\,{\rm ln}(q-1)$, $\exp(t)=(q-1)^{\alpha}$, the expression in the brackets (~~) at the right-hand side of (\ref{e18}) becomes
\begin{eqnarray} \label{e105}
& \mbox{} & \hspace*{-8mm}\frac{t\,e^t}{(e^t+q-1)^2}-\frac{\tau-3}{e^t+q-1}+(\tau-2)(\tau-3)\, \il_1^{\infty}\,\frac{w^{-\tau+1}\,dw}{e^{tw}+q-1} \nonumber \\[3.5mm]
& & \hspace*{-8mm}=~\frac{\alpha(q-1)^{\alpha-2}\,{\rm ln}(q-1)}{((q-1)^{\alpha-1}+1)^2} \nonumber \\[3.5mm]
& & \hspace*{-3mm}-~ \frac{\tau-3}{q-1}\,\left(\frac{1}{(q-1)^{\alpha-1}+1}-(\tau-2)\,\il_1^{\infty}\, \frac{w^{-\tau+1}\,dw}{(q-1)^{\alpha w-1}+1}\right)~.
\end{eqnarray}
As to the integral, we have
\beq \label{e106}
\lim_{q\pr\infty}\,\frac{1}{(q-1)^{\alpha w-1}+1}=1\,,~~1<w<\frac{1}{\alpha}\,;~~
\lim_{q\pr\infty}\,\frac{1}{(q-1)^{\alpha w-1}+1}=0\,,~~w>\frac{1}{\alpha}~.
\eq
Hence, by dominated convergence,
\beq \label{e107}
\lim_{q\pr\infty}\,\il_1^{\infty}\,\frac{w^{-\tau+1}\,dw}{(q-1)^{\alpha w-1}+1}= \il_1^{1/\alpha}\,w^{-\tau+1}\,dw~.
\eq
Since $\alpha<1$, we have that $(q-1)^{\alpha-1}\pr0$ as $q\pr\infty$, and so
\begin{eqnarray} \label{e108}
& \mbox{} & \lim_{q\pr\infty}\,\left(\frac{1}{(q-1)^{\alpha-1}+1}-(\tau-2)\,\il_1^{\infty}\,\frac{ w^{-\tau+1}\,dw}{(q-1)^{\alpha w-1}+1}\right) \nonumber \\[3.5mm]
& & =~1-(\tau-2)\,\il_1^{1/\alpha}\,w^{-\tau+1}\,dw=(\tau-2)\,\il_{1/\alpha}^\infty\, w^{-\tau+1}\,dw~.
\end{eqnarray}
Furthermore, since $\alpha\in(0,1)$,
\beq \label{e109}
\frac{\alpha(q-1)^{\alpha-2}\,{\rm ln}(q-1)}{((q-1)^{\alpha-1}+1)^2}= O((q-1)^{\alpha-2}\,{\rm ln}(q-1))=o\Bigl(\frac{1}{q-1}\Bigr)~,~~~~~~q\pr\infty~.
\eq
Therefore, from (\ref{e108}) and (\ref{e109}) we have that the expression in (\ref{e105}) is negative when $q$ is sufficiently large, i.e., $\cak''(\alpha\,{\rm ln}(q-1))<0$ when $q$ is sufficiently large.

\section{Limiting behaviour of $t_c''$ as $q\downarrow 2$} \label{sec7}
\mbox{} \\[-9mm]

The characterization of $t_c''$ as the unique positive zero $t=t_{\ast}$ of $\caf_0''(t)$ gives a convenient handle to study the behaviour of $t_c''$ as $q\downarrow 2$; such a characterization is not available for $t_c$ or $t_c'$. By \cite{ref1}, Sections~7.2 and 7.3, we have for $\tau\geq4$ that $t_{\ast}$ is the unique positive zero $t$ of the function
\beq \label{e110}
\Phi(t)=\il_t^{\infty}\,x^{-\tau+3}\,a(x)\,dx~,
\eq
where
\beq \label{e111}
a(x)=\frac{(q-1)\,e^x-e^{2x}}{(e^x+q-1)^3}~,~~~~~~x\geq0~.
\eq
Note that $a(x)$ has exponential decay as $x\pr\infty$, and that for any $x\geq0$
\beq \label{e112}
0\leq x<b\Rightarrow a(x)>0~;~~~~~~a(b)=0~;~~~~~~x>b\Rightarrow a(x)<0~,
\eq
where
\beq \label{e113}
0<b={\rm ln}(q-1)=q-2+O((q-2)^2)~.
\eq
Hence, $\Phi(t)<0$ for $t\geq b$, and so $t_c''=t_{\ast}<b={\rm ln}(q-1)$ since $\Phi(t_{\ast})=0$.

We write for $t\in(0,b)$
\beq \label{e114}
\Phi(t)=\il_t^b\,x^{-\tau+3}\,a(x)\,dx+\il_b^{\infty}\,x^{-\tau+3}\,a(x)\,dx=: T_1-T_2~,
\eq
where $T_1$ and $T_2$ are positive. We consider this for $b\downarrow 0$, and we should find $t$ such that $T_1=T_2$. We shall solve the equation $T_1=T_2$ approximately by finding the leading-order behaviour of $T_1$ and $T_2$ as $0<t\leq b$, $b\downarrow 0$. As to $T_1$, we use for this
\beq \label{e115}
a(x)=\tfrac18\,(b-x)+O(b^2)~,~~~~~~0\leq x\leq b~,
\eq
and we therefore approximate $T_1$ by
\beq \label{e116}
\tfrac18\,\il_t^b\,(b-x)\,x^{-\tau+3}\,dx=\tilde{T}_1~.
\eq
As to $T_2$, we have
\begin{eqnarray} \label{e117}
T_2 & = & \il_{{\rm ln}(q-1)}^{\infty}\, \frac{e^{2x}-(q-1)\,e^x}{(e^x+q-1)^3}\, x^{-\tau+3}\,dx=\il_{q-1}^{\infty}\,\frac{s-(q-1)}{(s+q-1)^3}\, ({\rm ln}\,s)^{-\tau+3}\,ds \nonumber \\[3.5mm]
& = & \il_0^{\infty}\,\frac{v}{(v+2(q-1))^3}\,({\rm ln}(v+q-1))^{-\tau+3}\,dv \nonumber \\[3.5mm]
& \approx & \il_0^{\infty}\,\frac{v}{(v+2)^3}\,({\rm ln}(1+v+b))^{-\tau+3}\,dv =\tilde{T}_2~,
\end{eqnarray}
where we have used the substitutions $s=e^x\geq q-1$ and $v=s-(q-1)\geq0$, together with (\ref{e113}).

We first consider $4\leq\tau<5$. Then we have
\beq \label{e118}
\lim_{b\downarrow0}\,\tilde{T}_2=\il_0^{\infty}\,\frac{v({\rm ln}(1+v))^{-\tau+3}} {(v+2)^3}\,dv=: C(\tau)~,
\eq
where $C(\tau)$ is finite and positive. For $\tilde{T}_1$ we set $t=\alpha b$ with $\alpha\in(0,1]$, and we have by the substitution $y=x/b\in[\alpha,1]$
\beq \label{e119}
\tilde{T}_1=\tfrac18\,b^{-\tau+5}\,\il_{\alpha}^1\,\frac{1-y}{y^{\tau-3}}\,dy~.
\eq
When $\alpha$ would be bounded away from 0 as $b\downarrow0$, we would have $\tilde{T}_1=O(b^{-\tau+5})\pr0$ as $b\downarrow0$ while $\tilde{T}_2\pr C(\tau)>0$ as $b\downarrow 0$. Therefore, we have $\alpha\pr0$ when we equate $\tilde{T}_1$ and $\tilde{T}_2$ as $b\downarrow0$, and then we have
\beq \label{e120}
\tilde{T}_1=\tfrac18\,b^{-\tau+5}\,\il_{\alpha}^1\,\frac{dy}{y^{\tau-3}}+ O(b^{-\tau+5})~.
\eq

In the case that $\tau=4$, we have
\beq \label{e121}
\tilde{T}_1=\frac{b}{8}\,{\rm ln}\Bigl(\frac{1}{\alpha}\Bigr)+O(b)~,
\eq
and equating this to $\tilde{T}_2$, we get from (\ref{e118}) the limiting behaviour
\beq \label{e122}
\alpha=\exp({-}8\,C(4)/b)~,~~~~~~t=b\alpha=b\,\exp({-}8\,C(4)/b)~,
\eq
showing exponential decay of $\alpha$ and $t$ as $b\downarrow0$.

In the case that $4<\tau<5$, we have
\beq \label{e123}
\tilde{T}_1=\tfrac18\,b^{-\tau+5}\,\frac{(1/\alpha)^{\tau-4}}{\tau-4}+ O(b^{-\tau+5})~,
\eq
and equating this to $\tilde{T}_2$, we get from (\ref{e118}) the limiting behaviour
\beq \label{e124}
\alpha=(8(\tau-4)\,C(\tau))^{\frac{-1}{\tau-4}}\,b^{\frac{5-\tau}{\tau-4}}~, ~~~~~~t=b\alpha=\Bigl(\frac{b}{8(\tau-4)\,C(\tau)}\Bigr)^{\frac{1}{\tau-4}}
\eq
as $b\downarrow0$.

Next, we consider the case $\tau=5$. Then we have from (\ref{e120})
\beq \label{e125}
\tilde{T}_1=\tfrac18\,\il_{\alpha}^1\,\frac{dy}{y^2}+O(1)= \frac{1}{8\alpha}+ O(1)~.
\eq
From (\ref{e117}) and ${\rm ln}(1+v+b)=(v+b)(1+O(v+b))$, we have
\begin{eqnarray} \label{e126}
\tilde{T}_2 & = & \il_0^1\,\frac{v}{(v+2)^3}~\frac{1}{(v+b)^2}\,dv+O(1) \nonumber \\[3.5mm]
& = & \tfrac18\,\il_0^1\,\frac{v\,dv}{(v+b)^2}+O(1)=\tfrac18\,{\rm ln}\Bigl(\frac1b\Bigr)+O(1)~.
\end{eqnarray}
Equating $\tilde{T}_1$ and $\tilde{T}_2$, we get the limiting behaviour
\beq \label{e127}
\alpha=\frac{1}{{\rm ln}\Bigl(\frac1b\Bigr)}~,~~~~~~t=b\alpha= \frac{b}{{\rm ln}\Bigl(\frac1b\Bigr)}
\eq
as $b\downarrow0$.

We finally consider the case $\tau>5$. From (\ref{e117}) we have
\beq \label{e128}
\tilde{T}_2=\tfrac18\,\il_0^1\,v\Bigl(\frac{1}{v+b}\Bigr)^{\tau-3}\,dv+ O\left[\il_0^1\,v\Bigl(\frac{1}{v+b}\Bigr)^{\tau-4}\,dv\right]~.
\eq
By partial integration
\begin{eqnarray} \label{e129}
\il_0^1\,v\Bigl(\frac{1}{v+b}\Bigr)^{\tau-3}\,dv & = & \left.\frac{v(v+b)^{-\tau+4}}{-\tau+4}\right|_0^1+\il_0^1\,\frac{(v+b)^{-\tau+4}}{\tau-4} \,dv \nonumber \\[3mm]
& = & \frac{b^{-\tau+5}}{(\tau-4)(\tau-5)}+O(1)~.
\end{eqnarray}
In a similar fashion
\beq \label{e130}
\il_0^1\,v\Bigl(\frac{1}{v+b}\Bigr)^{\tau-4}\,dv=O\left(\il_0^1\,(v+b)^{-\tau+5} \,dv\right)
\eq
in which the integral on the right-hand side is $O(1)$ when $5<\tau<6$, $O({\rm ln}(\frac1b))$ when $\tau=6$ and $O(b^{-\tau+6})$ when $\tau>6$. Thus, in all cases where $\tau>5$, we have
\beq \label{e131}
\tilde{T}_2=\frac{b^{-\tau+5}}{8(\tau-4)(\tau-5)}\,(1+o(1))~,~~~~~~ b\downarrow0~.
\eq
As to $\tilde{T}_1$, we now return to (\ref{e119}), and equating $\tilde{T}_2$ from (\ref{e131}) and $\tilde{T}_1$ from (\ref{e119}), we get for $\alpha\in(0,1)$ the equation
\beq \label{e132}
\il_{\alpha}^1\,\frac{1-y}{y^{\tau-3}}\,dy=\frac{1}{(\tau-4)(\tau-5)}~.
\eq
Now we have
\begin{eqnarray} \label{e133}
\il_{\alpha}^1\,\frac{1-y}{y^{\tau-3}}\,dy & = & \left.\Bigl(\frac{y^{-\tau+4}} {-\tau+4}-\frac{y^{-\tau+5}}{-\tau+5}\Bigr)\right|_{\alpha}^1 \nonumber \\[3mm]
& = & \frac{1}{(\tau-4)(\tau-5)}+\frac{\alpha^{-\tau+4}}{\tau-4}- \frac{\alpha^{-\tau+5}}{\tau-5}~.
\end{eqnarray}
Therefore, (\ref{e132}) can be solved in closed form for $\alpha$, and we get the limiting behaviour, as $b\downarrow0$,
\beq \label{e134}
\alpha=\frac{\tau-5}{\tau-4}~,~~~~~~t=b\alpha=\frac{\tau-5}{\tau-4}\,b= \frac{\tau-5}{\tau-4}\,{\rm ln}(q-1)~.
\eq

As said, we do not have a characerization of $t_c$ or $t_c'$ of the type that we have of $t_c''$ via (\ref{e110}), (\ref{e111}). Nevertheless, from a numerical investigation, solving $\cak(t)=0$, $\cak'(t)=0$ using the representations (\ref{e16}), (\ref{e17}), a picture, similar to the one for $t_c''$ as $b\downarrow0$, arises for $t_c$ and $t_c'$ as $b\downarrow0$.
\newpage
\section{The homogeneous case as limiting Pareto case with $\tau\pr\infty$} \label{sec8}
\mbox{} \\[-9mm]

We now consider the case that the Pareto pdf $f_W(w)$ in (\ref{e15}) is replaced by $\delta(w-1)$, also see \cite{ref1}, Appendix~C. This is the homogeneous case that arises from the Pareto case when we take $\tau\pr\infty$. The $\caf_0$ of (\ref{e1}) and the $\cak$ of (\ref{e6}) for the homogeneous case are given by
\beq \label{e135}
\caf_{0,H}(t)=\frac{e^t-1}{e^t+q-1} =1-\frac{q}{e^t+q-1}~,~~~~~~t\geq0~,
\eq
and
\beq \label{e136}
\cak_H(t)={\rm ln}\Bigl(\frac{e^t+q-1}{q}\Bigr)-\frac{q+1}{2q}\,t+\tfrac12\:\frac{q-1} {e^t+q-1}\,t~,~~~~~~t\geq0~,
\eq
respectively. By straightforward differentiation of (\ref{e136}), we get
\beq \label{e137}
\cak_H'(t)=\tfrac12\,(q-1)\Bigl(\frac1q-\frac{1}{e^t+q-1}-\frac{t\,e^t}{(e^t+q-1)^2} \Bigr)~,~~~~~~ t\geq0~,
\eq
and
\beq \label{e138}
\cak_H''(t)={-}\,\tfrac12\,(q-1)\,t\,\frac{(q-1)\,e^t-e^{2t}}{(e^t+q-1)^3}~,~~~~~~t\geq0~.
\eq
It is easy to show that $\cak_H(t)$ vanishes at $t=2\,{\rm ln}(q-1)$. Furthermore, it is seen at once that $\cak_H'(t)$ vanishes at $t=T$, with $t=T$ the unique positive solution of the equation (\ref{e25}). Finally, we have that $\cak_H''(t)$ vanishes at $t={\rm ln}(q-1)$. Therefore, the inequalities $t_c\leq2\,{\rm ln}(q-1)$, $t_c'\leq T$ and $t_c''\leq{\rm ln}(q-1)$, that hold for the Pareto case with finite $\tau$, are sharp in the sense that there is equality in the limit case $\tau\pr\infty$.

Observe that from (\ref{e16}), (\ref{e17}) and (\ref{e18}) we have
\begin{align}\label{e139}
\frac{\tau-1}{\tau-2}\cak(t)-\frac{t}{\tau-2}\cak'(t)&=\cak_H(t)+\frac{t}{(\tau-2)^2}~,\nonumber\\
\cak'(t)-\frac{t}{\tau-2}\cak''(t)&=\cak_H'(t)~ .
\end{align}

\section{Analysis of $T=t_{H,c}'$} \label{sec9}
\mbox{} \\[-9mm]

Denoting by $t_{H,c}$, $t_{H,c}'$ and $t_{H,c}''$ the unique positive zeros of $\cak_H(t)$, $\cak_H'(t)$ and $\cak_H''(t)$, respectively, with $\cak_H$ the $\cak$-function for the homogeneous case, we have seen in Section~\ref{sec8} that $t_{H,c}$ and $t_{H,c}''$ have an elementary closed form. Such a thing does not hold for $t_{H,c}'$ since the unique positive solution $T=t_{H,c}'$ of the equation (\ref{e25}) does not have a closed form. We shall, therefore, present a further analysis of $T$.

We shall show that $T(q)$ is an increasing function of $q \geq2$, with $T(2)=0$ and $T(\infty)=\infty$, and we shall show that $T(q)/{\rm ln}(q-1)$ decreases from $3/2$ at $q=2$ to 1 at $q=\infty$. Furthermore, we shall show that
\beq \label{e140}
T(q)=\tfrac32\,(q-2-\tfrac12\,(q-2)^2+\tfrac{13}{40}\,(q-2)^3+O((q-2)^4))~,~~~~~~ q\downarrow 2~,
\eq
where we observe that
\beq \label{e141}
{\rm ln}(q-1)=q-2-\tfrac12\,(q-2)^2+\tfrac{13}{39}\,(q-2)^3+O((q-2)^4)~,~~~~~~ q\downarrow 2~.
\eq
This agrees with the upper bound $T(q)<\frac32\,{\rm ln}(q-1)$ in (\ref{e76}). In addition, we shall show that
\beq \label{e142}
T(q)={\rm ln}\,q+{\rm ln}({\rm ln}\,B)+O\Bigl(\frac{{\rm ln}({\rm ln}\,B)}{{\rm ln}\,B} \Bigr)~,~~~~~B=\frac{q}{e}~,~~~~~q\pr\infty~.
\eq
and this agrees asymptotically with the upper bound $T(q)\geq{\rm ln}(q-1)$ in (\ref{e76}).

The equation (\ref{e25}) for $t=T=T(q)$ can be written as
\beq \label{e143}
\frac1q\,(e^t+q-1)^2-t\,e^t-(e^t+q-1)=0
\eq
and this can be worked out further to
\beq \label{e144}
\frac1q\,(e^t-1)^2-(t-1)\,e^t-1=0~.
\eq
The quantity at the left-hand side of (\ref{e144}) decreases in $q\geq2$ for all $t>0$, it vanishes uniquely at $t=T(q)$, and is negative for $0<t<T(q)$ and positive for $t>T(q)$. It follows that $T(q)$ is an increasing function of $q\geq2$.

The proof that $T(q)/{\rm ln}(q-1)$ decreases in $q\geq2$ is considerably more complicated. With $\alpha\in(1,3/2)$, we consider
\beq \label{e145}
V(x,\alpha)=\frac1q\,(e^t-1)^2-(t-1)\,e^t-1=\frac{(x^{\alpha}-1)^2}{x+1}- (a\,{\rm ln}\,x-1)\,x^{\alpha}-1~,
\eq
where we have set
\beq \label{e146}
x=q-1\geq1~,~~~~~~t=\alpha\,{\rm ln}(q-1)~,
\eq
so that $x^{\alpha}=e^t$ and $t=\alpha\,{\rm ln}\,x$. Our goal is to show that $V(x,\alpha)$ has a unique zero $x(\alpha)>1$ and that $x(\alpha)$ decreases strictly in $\alpha\in(1,3/2)$, Once this goal is reached, we have from $V(x(\alpha),\alpha)=0$ and (\ref{e146}) that
\beq \label{e147}
T(q(\alpha))=\alpha\,{\rm ln}\,x(\alpha)~,~~~~~~q(\alpha)=x(\alpha)+1~.
\eq
Denoting by $\alpha(q)$, $q>2$, the inverse function of $q(\alpha)$, $\alpha\in(1,3/2)$, we then have that
\beq \label{148}
\frac{T(q)}{{\rm ln}(q-1)}=\alpha(q)
\eq
is a decreasing function of $q>2$.

To show existence and uniqueness of a zero $x>1$ of $V(x,\alpha)$ when $\alpha\in(1,3/2)$, we write
\beq \label{e149}
V(x,\alpha)=\frac{(y-1)^2}{y^{\beta}+1}-y({\rm ln}\,y-1)-1~,
\eq
where we have set
\beq \label{e150}
\beta=\frac{1}{\alpha}\in(\tfrac23,1)~,~~~~~~y=x^{\alpha}\geq1~.
\eq
With an argument quite similar to the proof of (\ref{e74}), we have
\beq \label{e151}
\frac{(y-1)^2}{y+1}-y({\rm ln}\,y-1)-1<0<\frac{(y-1)^2}{y^{2/3}+1}-y({\rm ln}\,y-1)-1~, ~~~~~~y>1~,
\eq
so that the right-hand side of (\ref{e149}) is zero-free in $y>1$ when $\beta\geq1$ or $\beta\leq 2/3$. Hence, we concentrate on the case $\beta\in(2/3,1)$.

\begin{lem} \label{lem9.1}
We have for $y-1$ small positive
\beq \label{e152}
\frac{(y-1)^2}{y^{\beta}+1}-y({\rm ln}\,y-1)-1<0~,
\eq
and for $y-1$ large positive
\beq \label{e153}
\frac{(y-1)^2}{y^{\beta}+1}-y({\rm ln}\,y-1)-1>0~.
\eq
\end{lem}

\noindent
{\bf Proof.}~~The left-hand side of (\ref{e153}) tends to $\infty$ as $y\pr\infty$ since $\beta<1$, and is therefore positive when $y$ is large.

As to (\ref{e152}), we write $y=1+t$ with $t$ small positive. Then
\beq \label{e154}
\frac{(y-1)^2}{y^{\beta}+1}-y({\rm ln}\,y-1)-1=\frac{t^2}{(1+t)^{\beta}+1}- ((1+t)\,{\rm ln}(1+t)-t)~.
\eq
We have
\begin{eqnarray} \label{e155}
\frac{t^2}{(1+t)^{\beta}+1} & = & \frac{t^2}{2+\beta t+\tfrac12\,\beta(\beta-1)\,t^2+...} \nonumber \\[3mm]
& = & \tfrac12\,t^2(1+\tfrac12\,\beta t+\tfrac14\,\beta t^2+...)~.
\end{eqnarray}
At the same time, we have
\beq \label{e156}
(1+t)\,{\rm ln}(1+t)=\tfrac12\,t^2(1-\frac13\,t+\tfrac16\,t^2-...)~.
\eq
Hence,
\begin{eqnarray} \label{e157}
& \mbox{} & \frac{t^2}{(1+t)^{\beta}+1}-((1+t)\,{\rm ln}(1+t)-t) \nonumber \\[3mm]
& & =~\tfrac12\,t^2(\tfrac12\,(\tfrac23-\beta)\,t+\tfrac14\,(\beta-\tfrac23)\,t^2+...)<0
\end{eqnarray}
for small positive $t$ since $\beta>2/3$. This completes the proof. \\
\mbox{}

We conclude from Lemma~\ref{lem9.1} that the right-hand side of (\ref{e149}) has at least one zero $y>1$. As to uniqueness of this zero, we write
\begin{eqnarray} \label{e158}
& \mbox{} & \frac{(y-1)^2}{y^{\beta}+1}-y({\rm ln}\,y-1)-1 \nonumber \\[3mm]
& & =~\frac{1}{y^{\beta}+1}\,(y^2-2y-y(y^{\beta}+1)\,{\rm ln}\,y+y(y^{\beta}+1)-y^{\beta}) \nonumber \\[3mm]
& & =~\frac{1}{y^{\beta}+1}\,(y^2-y-(y^{\beta+1}+y)\,{\rm ln}\,y+y^{\beta+1}- y^{\beta}) \nonumber \\[3mm]
& & =~\frac{y^{\beta}}{y^{\beta}+1}\,p(y)~,
\end{eqnarray}
where
\beq \label{e159}
p(y)=y^{2-\beta}-y^{1-\beta}-(y+y^{1-\beta})\,{\rm ln}\,y+y-1~,~~~~~~ y\geq1~.
\eq

\begin{lem} \label{lem9.2}
We have $p(1)=0$, $p(y)<0$ for small positive $y-1$, and $p(y)>0$ for large positive $y-1$. Furthermore, $p(y)$, $y>1$, has a unique inflection point, that is, there is a unique $\hat{y}>1$ such that $p(y)$ is concave in $y\in(1,\hat{y})$ and convex in $y\in(\hat{y},\infty)$.
\end{lem}

\noindent
{\bf Proof.}~~The first statement follows from Lemma~\ref{lem9.1}. As to the second statement, we compute
\begin{eqnarray} \label{e160}
p''(y) & \!\!= & \!\!((2-\beta)\,y^{1-\beta}-(2-\beta)\,y^{-\beta}-(1+(1-\beta)\,y^{-\beta})\,{\rm ln}\,y)' \nonumber \\[3mm]
& \!\!= & \!\!(2-\beta)(1-\beta)\,y^{-\beta}-(\beta^2-3\beta+1)\,y^{-\beta-1}+ (1-\beta)\,\beta\,y^{-\beta-1}\,{\rm ln}\,y-\frac1y \nonumber \\[3mm]
& \!\!= & \!\!y^{-\beta-1}\,r(y)~,
\end{eqnarray}
where
\beq \label{e161}
r(y)=(2-\beta)(1-\beta)\,y-(\beta^2-3\beta+1)+(1-\beta)\,\beta\,{\rm ln}\,y-y^{\beta} ~,~~~~~~y\geq1~.
\eq
We have 
\beq \label{e162}
p(1)=p'(1)=p''(1)=0=r(1)~.
\eq
Furthermore,
\beq \label{e163}
r'(y)=(2-\beta)(1-\beta)+(1-\beta)\,\beta\,\frac1y-\beta\,y^{\beta-1}~; ~~~~~~r'(1)=2-3\beta<0~,
\eq
since $\beta>2/3$. Finally,
\beq \label{e164}
r''(y)=(1-\beta)\,\beta\Bigl(\frac{1}{y^{2-\beta}}-\frac{1}{y^2}\Bigr)>0\,,~~y>1~;~~~~~~ r''(1)=0
\eq
since $\beta<1$. Therefore, $r(y)$ is strictly convex in $y\geq1$. From (\ref{e161}) we see that $r(y)>0$ for large $y>1$ since $\beta<1$, and from (\ref{e162}) and (\ref{e163}) we see that $r(y)<0$ for small positive $y-1$. Hence, $r(y)$ has at least one zero $>\,1$, and this zero is unique by strict convexity of $r(y)$. Denote this zero by $\hat{y}$.

We conclude from (\ref{e160}) that $p(y)$ is strictly concave in $y\in(1,\hat{y})$ and strictly convex in $y\in(\hat{y},\infty)$. This completes the proof. \\
\mbox{}

We conclude from Lemma~\ref{lem9.2} that $p(y)$ has precisely one zero in $(\hat{y},\infty)$. According to (\ref{e149}) and (\ref{e158}), we see that $V(x,\alpha)$ has precisely one zero $x(\alpha)$ in $(1,\infty)$ for any $\alpha\in(1,3/2)$.

We next show that $x(\alpha)$ decreases strictly in $\alpha\in(1,3/2)$. It follows from Lemma~\ref{lem9.1} that $V(x,\alpha)<0$ for small positive $x-1$ and that $V(x,\alpha)>0$ for large positive $x-1$. Hence we have $V(x,\alpha)<0$ for $x\in(1,x(\alpha))$ and $V(x,\alpha)>0$ for $x\in(x(\alpha),\infty)$, while $V(x(\alpha),\alpha)=0$.

\begin{lem} \label{lem9.3}
We have for $\alpha\in(1,3/2)$ that
\beq \label{e165}
\frac{\partial V}{\partial\alpha}\,(x,\alpha)>~0~,~~~~~~x\geq x(\alpha)~.
\eq
\end{lem}

\noindent
{\bf Proof.}~~We have from (\ref{e145})
\begin{eqnarray} \label{e166}
\frac{\partial V}{\partial \alpha} & = & \frac{2}{x+1}\,(x^{\alpha}-1)\,{\rm ln}\,x -x^{\alpha}\,{\rm ln}\,x-(\alpha\,{\rm ln} x-1)\,x^{\alpha}\,{\rm ln}\,x \nonumber \\[3mm]
& = & x^{\alpha}\,{\rm ln}\,x\Bigl(2\,\frac{x^{\alpha}-1}{x+1}-\alpha\,{\rm ln}\,x\Bigr)~.
\end{eqnarray}
Now assume that $x\geq x(\alpha)$ so that $V(x,\alpha)\geq0$, i.e.,
\beq \label{e167}
\frac{(x^{\alpha}-1)^2}{x+1}-(\alpha\,{\rm ln}\,x-1)\,x^{\alpha}-1\geq0~.
\eq
Therefore,
\beq \label{e168}
\frac{x^{\alpha}-1}{x+1}\geq\frac{(a\,{\rm ln}\,x-1)\,x^{\alpha}+1}{x^{\alpha}-1}= \frac{x^{\alpha}}{x^{\alpha}-1}\,\alpha\,{\rm ln}\,x-1~.
\eq
Hence $\tfrac{\partial V}{\partial\alpha}\,(x,\alpha)>0$ when
\beq \label{e169}
2\,\frac{\xal}{\xal-1}\,\alpha\,{\rm ln}\,x-2>\alpha\,{\rm ln}\,x~,
\eq
i.e., when
\beq \label{e170}
\Bigl(2\,\frac{\xal}{\xal-1}-1\Bigr)\,\alpha\,{\rm ln}\,x=\frac{\xal+1}{\xal-1}\,\alpha\,{\rm ln}\,x>2~.
\eq
Hence, with $y=\xal>1$, we should show that $(y+1)\,{\rm ln}\,y/(y-1)>2$. This indeed holds, see below (\ref{e151}), and this completes the proof. \\
\mbox{}

Let $\alpha\in(1,3/2)$. We have from (\ref{e165}) that $V(x(\alpha),\alpha_1)>0$ when $\alpha_1-\alpha$ is small positive. Therefore $x(\alpha)>x(\alpha_1)$ when $\alpha_1-\alpha$ is small positive. That is, $x(\alpha)$ decreases in $\alpha\in(1,3/2)$. \\
\mbox{}

We next show the result (\ref{e140}) on the behaviour of $T(q)$ when $q-2\downarrow 0$. To this end, we use the functions $a_n(q)$, introduced in \cite{ref2}, Section~5, to expand
\beq \label{e171}
\frac{2}{q-1}\,\cak_H'(t)=\frac1q-\frac{1}{e^t+q-1}-\frac{t\,e^t}{(e^t+q-1)^2}~.
\eq
We have, by definition \cite{ref2}, (75),
\beq \label{e172}
\frac{1}{e^t+q-1}=\sum_{n=0}^{\infty}\,a_n(q)\,t^n~,~~~~~~|t|<\pi~.
\eq
Since
\beq \label{e173}
\frac{t\,e^t}{(e^t+q-1)^2}={-}t\,\frac{d}{dt}\,\Bigl(\frac{1}{e^t+q-1}\Bigr)={-} \,\sum_{n=1}^{\infty}\,n\,a_n(q)\,t^n~,~~~~~|t|<\pi~,
\eq
we find (using $a_0(q)=1/q$)
\begin{eqnarray} \label{e174}
\frac{2}{q-1}\,\cak_H'(t) & = & \frac1q-\Bigl(\frac1q+\sum_{n=1}^{\infty}\,a_n(q)\,t^n\Bigr) +\sum_{n=1}^{\infty}\,n\,a_n(q)\,t^n \nonumber \\[3.5mm]
& = & \sum_{n=2}^{\infty}\,(n-1)\,a_n(q)\,t^n~,~~~~~~|t|<\pi~.
\end{eqnarray}

We proceed as in \cite{ref2}, Section~6, where the unique zero $t_c(q)$ of the $\cak$-function corresponding to the exponential pdf $\exp({-}w)\,\cax_{[0,\infty)}(w)$ has been approximated as $q-2\downarrow0$ using the functions $a_n(q)$. In the present case, the details are much simpler since no awkward truncation analysis is needed, the expansion in (\ref{e174}) being valid for $|t|<\pi$. We get for $t=T(q)$ from $\cak_H'(t)=0$ the equaton
\beq \label{e175}
a_2(q)\,t^2+2a_3(q)\,t^3+3a_4(q)\,t^4+4a_5(q)\,t^5+...=0~,
\eq
so that
\beq \label{e176}
t=\frac{-a_2(q)}{2a_3(q)}-\frac32~\frac{a_4(q)}{a_3(q)}\,t^2-2\,\frac{a_5(q)}{a_3(q)}\,t^3+O(t^4)~.
\eq
We have by \cite{ref2}, (92), (93) and (117), (164)
\beq \label{e177}
\frac{-a_2(q)}{2a_3(q)}= \frac{-3q(q-2)}{2q^2-12q+12}= 3(\tfrac12\,v-\tfrac14\,v^2+\tfrac12\,v^3+O(v^4))~,~~~~~v=q-2\downarrow 0~.
\eq
Furthermore, from \cite{ref2}, (77) and below, we have $a_2(2)=a_4(2)=...=0$. Using \cite{ref2}, (88),
\beq \label{e178}
a_n(q)+(q-1)\,a_n'(q)={-}(n+1)\,a_{n+1}(q)~,~~~~~~n=1,2,...~,
\eq
we then get
\beq \label{e179}
a_4(q)=a_4'(2)\,v+O(v^2)={-}5a_5(2)\,v+O(v^2)~,~~~~~~v=q-2~.
\eq
Finally, from \cite{ref2}, (93) and (95),
\beq \label{e180}
a_3(2)=\frac{1}{48}~,~~~~~~a_5(2)=\frac{-1}{480}~.
\eq
Then combining (\ref{e176}), (\ref{e177}), (\ref{e179}) and (\ref{e180}) and using that $t=T(q)=O({\rm ln}(q-1))=O(v)$, we get
\begin{eqnarray} \label{e181}
t & = & 3(\tfrac12\,v-\tfrac14\,v^2+\tfrac12\,v^3+O(v^4))-\tfrac34\, (v+O(v^2))\,t^2+(\tfrac15+O(v))\,t^3 \nonumber \\[3mm]
& = & \tfrac32\,v-\tfrac34\,v^2+\tfrac32\,v^3-\tfrac34\,v\,t^2+\tfrac15\,t^3+O(v^4)~, ~~~~~v=q-2\downarrow0~.
\end{eqnarray}
Using that $t=O(v)$, we first get $t=\frac32\,v+O(v^2)$ from (\ref{e181}), and inserting this into the second line of (\ref{e181}), we get
\begin{eqnarray} \label{e182}
t & = & \tfrac32\,v-\tfrac34\,v^2+\tfrac32\,v^3-\tfrac34\,v(\tfrac32\,v)^2+\tfrac15\, (\tfrac32\,v)^3+O(v^4) \nonumber \\[3mm]
& = & \tfrac32\,v-\tfrac34\,v^2+\tfrac{39}{80}\,v^3+O(v^4)~,~~~~~~v=q-2\downarrow0~,
\end{eqnarray}
and this is (\ref{e140}). \\
\mbox{}

We finally show the result (\ref{e142}) on the asymptotics of $T(q)$ as $q\pr\infty$. We start by rewriting the equation (\ref{e143}) for $t=T(q)$ as
\beq \label{e183}
\Bigl(\frac{e^t}{q}\Bigr)^2-(t-1)\,\frac{e^t}{q}-\frac2q~\frac{e^t}{q}-\frac1q\,\Bigl(1-\frac1q \Bigr)=0~.
\eq
This is achieved by working out $(e^t+q-1)^2$ as $e^{2t}+2(q-1)\,e^t+(q-1)^2$, collecting terms and division by $q$. Thus we get the equation
\beq \label{e184}
s^2-(t-1)\,s-\frac2q\,s-\frac1q\,\Bigl(1-\frac1q\Bigr)=0~,~~~~~~s=\frac{e^t}{q}~.
\eq
Divide by $s$, set $B=q/e$, so that
\beq \label{e185}
t-1={\rm ln}(qs)-1={\rm ln}\Bigl(\frac{q}{e}\,s\Bigr)={\rm ln}\,B+{\rm ln}\,s~, ~~~~~~B=q/e~,
\eq
and we get the equation
\beq \label{e186}
s-{\rm ln}\,s-\frac2q-\frac1q\,\Bigl(1-\frac1q\Bigr)\,\frac1s={\rm ln}\,B~,~~~~~~ B=q/e~.
\eq
The left-hand side of (\ref{e186}) is increasing in $s>1/2$ for any $q>2$.

We aim in (\ref{e142}) at a result for $t=T(q)={\rm ln}(q\,s)$ in which a term ${\rm ln}({\rm ln}\,B)$ appears. We want here ${\rm ln}({\rm ln}\,B)$ to be well-defined and positive, and thus we require
\beq \label{e187}
{\rm ln}\,B>1~,~~~~~~B>e~,~~~~~~q>e^2~.
\eq
The left-hand side of (\ref{e186}) increases in $s>1/2$ and is less than 1 at $s=1$. Since ${\rm ln}\,B>1$, we therefore restrict to $s>1$. Then ${\rm ln}\,s>0$ and so we have $s>{\rm ln}\,B$ with ${\rm ln}\,B\pr\infty$ as $q\pr\infty$.

We continue by considering, instead of (\ref{e186}), the equation 
\beq \label{e188}
s-{\rm ln}\,s={\rm ln}\,B~,~~~~~~{\rm ln}\,B>1~,~~~~~~s>1~.
\eq
Hence, we have deleted from (\ref{e186}) the term
\beq \label{e189}
\frac2q+\frac1q\,\Bigl(1-\frac1q\Bigr)\,\frac1s\in\Bigl(\frac2q\,,\,\frac3q\Bigr)~.
\eq
The function $s\geq1\mapsto s-{\rm ln}\,s$ is strictly increasing, and both equations (\ref{e186}) and (\ref{e188}) have unique solutions when ${\rm ln}\,B>1$. The solution of (\ref{e186}) exceeds the solution of (\ref{e188}) in which ${\rm ln}\,B$ is replaced by ${\rm ln}\,B+\tfrac2q$ and is less than the solution of (\ref{e188}) in which ${\rm ln}\,B$ is replaced by ${\rm ln}\,B+\tfrac3q$, see (\ref{e189}). These two solutions of (\ref{e188}) have absolute difference of the order $1/q$, since $(s-{\rm ln}\,s)'=1-1/s$, which is therefore exponentially small in ${\rm ln}\,B$.

With ${\rm ln}\,B>1$, the solution $s>1$ of (\ref{e188}) can be shown to satisfy
\beq \label{e190}
{\rm ln}\,B<s\leq\frac{e}{e-1}\,{\rm ln}\,B~.
\eq
The first inequality in (\ref{e190}) is obvious and for the second inequality we use convexity of the function $s\geq1\mapsto s-{\rm ln}\,s$, so that
\beq \label{e191}
{\rm ln}\,B=s-{\rm ln}\,s\geq e-1+\Bigl(1-\frac1e\Bigr)(s-e)~.
\eq

We shall now approximate the solution $s$ of (\ref{e188}) by iterating (\ref{e188}) a few times. Thus, we have
\begin{eqnarray} \label{e192}
s & = & {\rm ln}\,B+{\rm ln}\,s={\rm ln}\,B+{\rm ln}({\rm ln}\,s+{\rm ln}\,B) \nonumber \\[3mm]
& = & {\rm ln}\,B+{\rm ln}({\rm ln}\,B)+{\rm ln}\Bigl(1+\frac{{\rm ln}\,s}{{\rm ln}\,B}\Bigr)~.
\end{eqnarray}
Now from (\ref{e190}),
\beq \label{e193}
\frac{{\rm ln}({\rm ln}\,B)}{{\rm ln}\,B}<\frac{{\rm ln}\,s}{{\rm ln}\,B}\leq \frac{{\rm ln}({\rm ln}\,B)+{\rm ln}\Bigl(\dfrac{e}{e-1}\Bigr)}{{\rm ln}\,B}~.
\eq
Hence,
\beq \label{e194}
s={\rm ln}\,B+{\rm ln}({\rm ln}\,B)+O\Bigl(\frac{{\rm ln}({\rm ln}\,B)}{{\rm ln}\,B} \Bigr)~,
\eq
where the implicit constant in $O$ is of the order unity.

Recalling that $T={\rm ln}(qs)$, with exponentially small error in ${\rm ln}\,B$, we have
\begin{eqnarray} \label{e195}
T & = & {\rm ln}\,q+{\rm ln}\,s={\rm ln}\,q+{\rm ln}({\rm ln}\,B+{\rm ln}\,s) \nonumber \\[3mm]
& = & {\rm ln}\,q+{\rm ln}({\rm ln}\,B)+{\rm ln}\Bigl(1+\frac{{\rm ln}\,s}{{\rm ln}\,B}\Bigr) \nonumber \\[3mm]
& = & {\rm ln}\,q+{\rm ln}({\rm ln}\,B)+O\Bigl(\frac{{\rm ln}({\rm ln}\,B)} {{\rm ln}\,B}\Bigr)~,~~~~~~B=q/e~,
\end{eqnarray}
and this is (\ref{e142}).

\setcounter{equation}{0}
\renewcommand{\theequation}{A\arabic{equation}}
\section*{Appendix A. Proof of (\ref{e74})} \label{appA}
\mbox{} \\[-9mm]

We show that for $q>2$ 
\beq \label{a1}
\psi(q-1)>0>\psi((q-1)^{3/2})~,
\eq
where 
\beq \label{a2}
\psi(y)=y(1+{\rm ln}\,y)+q-1-\frac1q\,(y+q-1)^2~,~~~~~~y\geq1~.
\eq

As to the first inequality in (\ref{a1}), we let $x=q-1>1$, and we compute
\beq \label{a3}
\psi(q-1)=2x+x\,{\rm ln}\,x-\frac{4x^2}{1+x}=\frac{x}{1+x}\, ((x+1)\,{\rm ln}\,x-2x+2)~.
\eq
It is easy to show that $(x+1)\,{\rm ln}\,x-2x+2>0$ when $x>1$.

As to the second inequality in (\ref{a1}), we let $y=(q-1)^{3/2}$, $q=1+y^{2/3}$, with $q>2$, $y>1$. We compute
\begin{eqnarray} \label{a4}
\psi(y) & = & y(1+{\rm ln}\,y)+y^{2/3}-\frac{(y+y^{2/3})^2}{1+y^{2/3}} \nonumber \\[2.5mm]
& = & \frac{1}{1+y^{2/3}}\,(y(1+y^{2/3})(1+{\rm ln}\,y)+y^{2/3}(1-y^{4/3})-2y^{5/3})~.
\end{eqnarray}
Setting $y=v^3$, $v=y^{1/3}>1$, we shall show that
\beq \label{a5}
v^3(1+v^2)(1+3\,{\rm ln}\,v)+v^2(1-v^4)-2v^5<0~,~~~~~~v>1~.
\eq
Dividing through by $v^2$, we have that (\ref{a5}) holds when
\beq \label{a6}
\xi(v):=1+v-v^3-v^4+3v\,{\rm ln}\,v+3v^3\,{\rm ln}\,v<0~,~~~~~~v>1~.
\eq
We have $\xi(v)={-}\varp(v)$, see (\ref{e54}), and so (\ref{a6}) follows.

\setcounter{equation}{0}
\renewcommand{\theequation}{B\arabic{equation}}
\section*{Appendix B. Proof of rapid convergence of $t_c/\rln\,p$ to $2(\tau-2)/(\tau-1)$ as $p=q-1\pr\infty$} \label{appB}
\mbox{} \\[-9mm]

We show that
\beq \label{b1}
x_c=\frac{t_c}{\rln\,p}=L(\tau)(1+O(1/p^{R(\tau)}))~,~~~~~~p=q-1\pr\infty~,
\eq
where
\beq \label{b2}
L(t)=2\,\frac{\tau-2}{\tau-1}~,~~~~~~R(\tau)=L(\tau)-1=\frac{\tau-3}{\tau-1}~.
\eq
We have, when $\tau\geq4$, that $t_c$ is the unique positive zero of
\beq \label{b3}
\cak(t)=\frac{\tau{-}2}{\tau{-}1}\,\rln\Bigl(\frac{e^t+q-1}{q}\Bigr)+\Bigl( \frac{1}{\tau-2}-\frac{q+1}{2q}\Bigr)\,t+ \frac{(\tau-2)(\tau-3)}{2(\tau-1)}\,(q-1)\,t\,D(t)\,,
\eq
where
\beq \label{b4}
D(t)=\il_1^{\infty}\,\frac{w^{-\tau+1}\,dw}{e^{tw}+q-1}~,~~~~~~t\geq0~.
\eq
Write $t=x\,\rln\,p$, $t_c=x_c\,\rln\,p$ with $x,x_c>0$ and $p=q-1$. Then, (see (\ref{e91+1}), (\ref{e91+2})
\beq \label{b5}
\rln\Bigl(\frac{e^t+q-1}{q}\Bigr)=(x-1)\,\rln\,p+r(x)~,
\eq
where
\beq \label{b6}
r(x)=\rln\Bigl(\frac{1+1/p^{x-1}}{1+1/p}\Bigr)~,~~~~~~x>0~.
\eq
We have from $\cak(x_c\,\rln\,p)=0$ that
\begin{eqnarray} \label{b7}
0 & = & \frac{\tau-2}{\tau-1}\,((x-1)\,\rln\,p+r(x))+\Bigl(\frac{1}{\tau-2}-\frac{p+2}{2(p+1)}\Bigr) \,x\,\rln\,p \nonumber \\[3mm]
& & +~\frac{(\tau-2)(\tau-3)}{2(\tau-1)}\,p\,D(t)\,x\,\rln\,p \nonumber \\[3mm]
& = & \Bigl(\frac{\tau-2}{\tau-1}+\frac{1}{\tau-2}-\frac12-\frac{1}{2(p+1)}\Bigr) \,x\,\rln\,p \nonumber \\[3mm]
& & +~\frac{\tau-2}{\tau-1}\,r(x)-\frac{\tau-2}{\tau-1}\,\rln\,p +\frac{(\tau-2)(\tau-3)}{2(\tau-1)}\,p\,D(t)\,x\,\rln\,p \nonumber \\[3mm]
& = & \Bigl(\frac12\,\Bigl(1-\frac{1}{p+1}\Bigr)+\frac{(\tau-2)(\tau-3)}{2(\tau-1)} \,p\,D(t)\Bigr)\,x\,\rln\,p \nonumber \\[3mm]
& & -~\frac{\tau-2}{\tau-1}\,\Bigl(1-\frac{r(x)}{\rln\,p}\Bigr)\,\rln\,p~,
~~~~~~x=x_c\,,~~t=t_c~.
\end{eqnarray}
Hence, dividing through by $\frac12\,\rln\,p$, we get
\beq \label{b8}
x=2\,\frac{\tau-2}{\tau-1}~\frac{1-r(x)/\rln\,p}{1-1/(p+1)+(\tau-2)(\tau-3)\,p\,D(t)/(\tau-1)} ~,~~~~~~x=x_c\,,~~t=t_c~.
\eq
The next step in the proof of the result (\ref{b1}) consists in bounding $r(x)$ and $p\,D(t)$. We have, since $x=x_c<2(\tau-2)/(\tau-1)<2$,
\beq \label{b9}
0<r(x)<\frac{1}{p^{x-1}}~.
\eq
Furthermore, there is the following result. \\ \\
{\bf Lemma.}~~{\em We have}
\beq \label{b10}
\frac{1}{p^{x-1}+1}~\frac{1}{\tau-1+x\,\rln\,p}<p\,D(t)<\frac{1}{p^{x-1}}~\frac{1}{\tau-2+x\,\rln\,p} ~,
\eq
{\em where $t=x\,\rln\,p$, $x\geq0$, $p>1$ and $\tau\geq4$.} \\ \\
{\bf Proof.}~~We have
\begin{eqnarray} \label{b11}
\hspace*{-5mm}p\,D(t) & = & \il_1^{\infty}\,\frac{p\,w^{-\tau+1}\,dw}{e^{tw}+p}=\il_1^{\infty}\, \frac{w^{-\tau+1}\,dw}{p^{xw-1}+1} \nonumber \\[3.5mm]
& = & \il_1^{\infty}\,\exp({-}(\tau-1)\,\rln\,w)-\rln(p^{xw-1}+1)\,dw \nonumber \\[3.5mm]
& < & \il_1^{\infty}\,\exp({-}(\tau-1)\,\rln\,w-(xw-1)\,\rln\,p)\,dw \nonumber \\[3.5mm]
& = & \il_0^{\infty}\,\exp({-}(\tau-1)\,\rln(1+v)-(x(1+v)-1)\,\rln\,p)\,dv \nonumber \\[3.5mm]
& = & \frac{1}{p^{x+1}}\,\il_0^{\infty}\,\exp({-}(\tau-1)\,\rln(1+v)-xv\,\rln\,p)\,dv \nonumber \\[3.5mm]
& < & \frac{1}{p^{x-1}}\,\il_0^{\infty}\,\exp({-}(\tau-1)\,\rln(1+v)-x\,\rln\,p\,\rln(1+v)\,dv \nonumber \\[3.5mm]
& = & \frac{1}{p^{x-1}}\,\il_0^{\infty}\,(1+v)^{-(\tau-1)-x\,\rln\,p}\,dv= \frac{1}{p^{x-1}}~\frac{1}{\tau-2+x\,\rln\,p}~.
\end{eqnarray}
At the same time, we have
\begin{eqnarray} \label{b12}
p\,D(t) & = & \il_1^{\infty}\,\exp({-}(\tau-1)\,\rln\,w-\rln(p^{xw-1}))\, \frac{dw}{1+(1/p)^{xw-1}} \nonumber \\[3.5mm]
& > & \frac{1}{1+1/p^{x-1}}\,\il_1^{\infty}\,\exp({-}(\tau-1)\,\rln\,w-(xw-1)\,\rln\,p) \,dw \nonumber \\[3.5mm]
& = & \frac{1}{p^{x-1}+1}\,\il_0^{\infty}\,\exp({-}(\tau-1)\,\rln(1+v)-xv\,\rln\,p)\,dv \nonumber \\[3.5mm]
& > & \frac{1}{p^{x-1}+1}\,\il_0^{\infty}\,\exp({-}(\tau-1)\,v-xv\,\rln\,p)\,dv \nonumber \\[3.5mm]
& = & \frac{1}{p^{x-1}+1}~\frac{1}{\tau-1+x\,\rln\,p}~.
\end{eqnarray}
This completes the proof. \\
\mbox{}

Below we shall need only the upper bound in (\ref{b10}), but the lower bound is of interest as well sine it gives an idea of sharpness of the upper bound.

In principle, the eventual rapid convergence of $x=x_c$ to $2(\tau-2)/(\tau-1)$ follows from (\ref{b8}) and the first item in (\ref{e29}), so that $x_c\pr 2(\tau-2)/(\tau-1)>1$ when $p=q-1\pr\infty$. By (\ref{b9}) and (\ref{b10}), this means that the factor
\beq \label{b13}
\frac{1-r(x)/\rln\,p}{1-1/(p+1)+(\tau-2)(\tau-3)\,p\,D(t)/(\tau-1)}\pr1~,~~~~~~ x=x_c\,,~~t=t_c~,
\eq
see the right-hand side of (\ref{b8}), exponentially fast in $\rln\,p$ as $p\pr\infty$. Indeed, we have from (\ref{b9}), (\ref{b10}) and (\ref{b8})
\begin{eqnarray} \label{b14}
0 & < & 2\,\frac{\tau-2}{\tau-1}-x=2\,\frac{\tau-2}{\tau-1}~\frac {(\tau-1)(\tau-3)\,p\, D(t)+r(x)/\rln\,p-1/(p+1)} {1-1/(p+1)+(\tau-2)(\tau-3)\,p\,D(t)/(\tau-1)} \nonumber \\[3mm]
& < & 2\,\frac{\tau-2}{\tau-1}~\frac{(1+1/\rln\,p)/p^{x-1}-1/(p+1)} {1-1/(p+1)}< 2\,\frac{\tau-2}{\tau-1}~\frac{1}{p^{x-1}}\,\Bigl(1+\frac{1}{\rln\,p}\Bigr)~, \nonumber \\[3mm]
& & \hspace*{7.5cm}x=x_c\,,~~t=t_c~,
\end{eqnarray}
provided that $p$ is so large that $(1+1/\rln\,p)/p^{x-1}<1$.

With the tools developed in the present section, we can give more precise results on the convergence of $x_c$ to $2(\tau-2)/(\tau-1)$, and to this end we proceed as follows. We fix an $\eps\in(0,2(\tau-2)/(\tau-1)-1)=(0,(\tau-3)/(\tau-1))$, and we ask ourselves for which $p>1$ do we have $\cak(x\,\rln\,p)<0$ with $x=2(\tau-2)/(\tau-1)-\eps$; for such $p$ we have $x_c>x\,\rln\,p$. We have, see the computation in (\ref{b7}),
\begin{eqnarray} \label{b15}
& \mbox{} & \cak(x\,\rln\,p) \nonumber \\[3mm]
& & =~\tfrac12\,x\,\rln\,p-\frac{\tau-2}{\tau-1}\,\rln\,p+\frac{\tau-2}{\tau-1}\,r(x)+
\frac{(\tau-2)(\tau-3)} {2(\tau-1)}\cdot x\,\rln\,p\cdot p\,D(t) \nonumber \\[3mm]
& & \hspace*{6.5cm}-~\frac{x}{2(p+1)}\,\rln\,p \nonumber \\[3mm]
& & =~\tfrac12\,\rln\,p\Bigl(x-2\,\frac{\tau-2}{\tau-1}+2\,\frac{\tau-2}{\tau-1}\,r(x)+ \frac{(\tau-2)(\tau-3)}{\tau-1}\,x\,p\,D(t)-\frac{x}{p+1}\Bigr)~. \nonumber \\[2mm]
\mbox{}
\end{eqnarray}
Noting that $x-2(\tau-2)/(\tau-1)={-}\eps$, we see that $\cak(x\,\rln\,p)<0$ when
\beq \label{b16}
2\,\frac{\tau-2}{\tau-1}~\frac{r(x)}{\rln\,p}+ \frac{(\tau-2)(\tau-3)}{\tau-1}\,x\,p\,D(t)-\frac{x}{p+1}<\eps~.
\eq
By (\ref{b10}) we have that (\ref{b16}) holds when
\beq \label{b17}
2\,\frac{\tau-2}{\tau-1}~\frac{r(x)}{\rln\,p}+ \frac{(\tau-2)(\tau-3)}{\tau-1}~\frac{x}{p^{x-1}(\tau-2+x\,\rln\,p)}- \frac{x}{p+1}<\eps~.
\eq
We have $r(x)<1/p^{x-1}$ and upon deleting the term $-x/(p+1)$ at the left-hand side of  (\ref{b17}), we see that (\ref{b17}) holds when
\beq \label{b18}
\frac{2}{\rln\,p}+\frac{(\tau-3)\,x}{\tau-2+x\,\rln\,p}< \frac{\tau-1}{\tau-2}\,\eps \,p^{x-1}~.
\eq
This latter criterion is less complicated than (\ref{b17}) itself.

Obviously, since $x>1$ is fixed, there is a $P_1=P_1(\tau,\eps)$ such that (\ref{b17}) holds when $p>P_1$ and a $P_2=P_2(\tau,\eps)$ such that (\ref{b18}) holds when $p>P_2$. In the table below, we display for $\tau=4,6,11$ a valid $P_1$ and $P_2$ for the case that $\eps=0.1$ (this $\eps$ is allowed for all $\tau\geq4$ since $2(\tau-2)/(\tau-1)-1\geq1/3$ when $\tau\geq4$).
\begin{center}
\begin{tabular}{rcc}
\mc{1}{c}{$\tau$} & \mc{1}{c}{$P_1(\tau,0.1)$} & \mc{1}{c}{$P_2(\tau,0.1)$} \\[2mm]
4 & 120 & 206 \\
6 & \hspace*{2mm}23 & \hspace*{2mm}57 \\
11 & \hspace*{4mm}8 & \hspace*{2mm}39
\end{tabular}
\end{center}
{\bf Example.}~~Take $\tau=6$, so that $2(\tau-2)/(\tau-1)=1.6\,$. For $p=q-1=19$, we compute $x_c=1.52116...\,$. With $\eps=0.1$, we have $x=2(\tau-2)/(\tau-1)-\eps=1.50<x_c$. Hence, $\cak(x\,\rln\,p)<0$ already holds for $p=19$ while $P_1(6,0.1)=23$. \\
\mbox{}

Assuming the conjectured lower bound (\ref{e83}) to hold for $t_c$, we get much better results. We then have that $x_c>2(\tau-5)/(\tau-4)$ for all $\tau\geq5$ and all $p>1$. For a fixed $\eps\in(0,2(\tau-2)/(\tau-1))$ and all $p>0$ there then holds
\beq \label{b19}
x=2\,\frac{\tau-2}{\tau-1}-\eps<2\,\frac{\tau-5}{\tau-4}\Rightarrow \cak(x\,\rln\,p)<0~.
\eq
We have $2(\tau-2)/(\tau-1)-\eps<2(\tau-5)/(\tau-4)$ if and only if $(\tau-1)(\tau-4)>6/\eps$. Hence, when $(\tau-1)(\tau-4)>6/\eps$, we have $\cak(x\,\rln\,p)<0$, i.e., $x_c>2(\tau-2)/(\tau-1)-\eps$ for all $p>1$. For instance, for $\eps=0.1$, we have $(\tau-1)(\tau-4)>6/\eps$ when $\tau>5/2+\sqrt{259/4}=10.54673...\,$. 

We return to (\ref{b14}). Letting $L(\tau)=2(\tau-2)/(\tau-1)$, $R(t)=L(\tau)-1= (\tau-3)/(\tau-1)$, as in (\ref{b2}), we have for $p>e$ by (\ref{b14})
\beq \label{b20}
0<L(\tau)-x<\frac{2L(\tau)}{p^{x-1}}~,~~~~~~x=x_c~,
\eq
i.e.,
\beq \label{b21}
L(\tau)>x>L(\tau)\Bigl(1-\frac{2}{p^{x-1}}\Bigr)~,~~~~~~x=x_c~.
\eq
Let $\eps\in(0,R(\tau))$ be fixed. For $p>P_1(\tau,\eps)$, we have $x=x_c>L(\tau)-\eps$, and so for $p>P_1(\tau,\eps)$
\beq \label{b22}
x=x_c>L(\tau)\Bigl(1-\frac{2}{p^{L(\tau)-1-\eps}}\Bigr)= L(\tau)\Bigl(1-\frac{2}{p^{R(\tau)-\eps}}\Bigr)~.
\eq
Therefore
\beq \label{b23}
p^{x-1}>p^{L(\tau)(1-2/p^{R(\tau)-\eps})-1}= p^{R(\tau)}/ p^{2L(\tau)/p^{R(\tau)-\eps}}~, ~~~~~~x=x_c~.
\eq
Now
\beq \label{b24}
p^{2L(\tau)/p^{R(\tau)-\eps}}=\exp\Bigl(\frac{2L(\tau)}{p^{R(\tau)-\eps}}\Bigr) \pr1~,~~~~~~p\pr\infty~,
\eq
and we conclude that
\beq \label{b25}
1-\frac{2}{p^{x-1}}=1-\frac{2}{p^{R(\tau)}}\,(1+o(1))~,~~~~~~p\pr\infty\,,~~ x=x_c~.
\eq
This gives (\ref{b1}),via (\ref{b21}). \\
\mbox{}

In the following three tables, we display for $\tau=4,6,11$ and $p=q-1=5,9,19,99,999$ the value of $x_c$ and of $(L(\tau)-x_c)\,p^{R(\tau)}$, with $L(\tau)=2(\tau-2)/(\tau-1)$ and $R(\tau)=(\tau-3)/(\tau-1)$. Note that $x_c$ is bounded between $2(\tau-5)/(\tau-4)$ (conjectured) and $2(\tau-2)/(\tau-1)$. \\ \\
\os{$\tau=4$}~~$L(\tau)=4/3$, $R(\tau)=1/3$, $2(\tau-5)/(\tau-4)={-}\infty$
\begin{center}
\begin{tabular}{rcc}
\mc{1}{c}{$p$} & $x_c$ & $(4/3-x_c)\,p^{1/3}$ \\[2mm]
5 & 0.730932 & 1.030092 \\
9 & 0.925431 & 0.848471 \\
19 & 1.079553 & 0.677188 \\
99 & 1.236418 & 0.448337 \\
999 & 1.304076 & 0.292475
\end{tabular}
\end{center}  
\mbox{} \\[2mm]
\os{$\tau=6$}~~$L(\tau)=8/5$, $R(\tau)=3/5$, $2(\tau-5)/(\tau-4)=1$
\begin{center}
\begin{tabular}{rcc}
\mc{1}{c}{$p$} & $x_c$ & $(8/5-x_c)\,p^{3/5}$ \\[2mm]
5 & 1.408134 & 0.503941 \\
9 & 1.470619 & 0.483522 \\
19 & 1.521168 & 0.461300 \\
99 & 1.573186 & 0.422418 \\
999 & 1.594047 & 0.375383
\end{tabular}
\end{center}  
\mbox{} \\[2mm]
\os{$\tau=11$}~~$L(\tau)=9/5$, $R(\tau)=4/5$, $2(\tau-5)/(\tau-4)=12/7= 1.714285...$
\begin{center}
\begin{tabular}{rcc}
\mc{1}{c}{$p$} & $x_c$ & $(9/5-x_c)\,p^{4/5}$ \\[2mm]
5 & 1.753286 & 0.169287 \\
9 & 1.766910 & 0.191907 \\
19 & 1.779241 & 0.218882 \\
99 & 1.793039 & 0.274903 \\
999 & 1.798652 & 0.338321
\end{tabular}
\end{center}  
\end{document}